\documentclass[longauth]{aa}

\newcommand{\ch}[1]{{$\rm #1$}}
\newcommand{\hl}[1]{{{#1}}}
\newcommand{\hll}[1]{{{#1}}}
\usepackage{graphicx}
\usepackage{txfonts}
\usepackage[hidelinks,colorlinks=true,linkcolor=blue,anchorcolor=black,citecolor=blue,filecolor=black,menucolor=black,runcolor=black,urlcolor=black]{hyperref}
\usepackage{orcidlink}
\usepackage{ragged2e}
\usepackage{float}
\usepackage{rotating}
\usepackage{multirow}
\usepackage{placeins}
\usepackage{array}
\newcolumntype{P}[1]{>{\centering\arraybackslash}p{#1}}
\newcolumntype{M}[1]{>{\centering\arraybackslash}m{#1}}
\newcommand{\PreserveBackslash}[1]{\let\temp=\\#1\let\\=\temp}
\newcolumntype{C}[1]{>{\PreserveBackslash\centering}p{#1}}
\usepackage{lipsum}
\usepackage{afterpage}
\usepackage[most]{tcolorbox}
\newtcbox{\myboxi}[1][]{nobeforeafter,tcbox raise base,colframe=green!50!black,colback=green!50!black,height=8pt,valign=center,raster valign=center,
  box align=base,sharp corners,top=0pt,bottom=0pt,left=0pt,right=2pt,
  boxrule=0pt,boxsep=2.5pt,before upper=\strut,#1}
\newcommand{\mybox}[2][1.1ex]{\raisebox{#1}{\myboxi{#2}}}
        
\newtcbox{\myboxxi}[1][]{nobeforeafter,tcbox raise base,colframe=black!50!white,colback=black!10!white,height=8pt,valign=center,raster valign=center,
  box align=base,sharp corners,top=0pt,bottom=0pt,left=0pt,right=2pt,
  boxrule=0pt,boxsep=2.5pt,before upper=\strut,#1}
\newcommand{\myboxx}[2][1.1ex]{\raisebox{#1}{\myboxxi{#2}}}
        
\newtcbox{\myboxxxi}[1][]{nobeforeafter,tcbox raise base,colframe=yellow!85!black!,colback=yellow!85!black!,height=8pt,valign=center,raster valign=center,
  box align=base,sharp corners,top=0pt,bottom=0pt,left=0pt,right=2pt,
  boxrule=0pt,boxsep=2.5pt,before upper=\strut,#1}
\newcommand{\myboxxx}[2][1.1ex]{\raisebox{#1}{\myboxxxi{#2}}}
\begin{document}

   \title{MINDS: The very low-mass star and brown dwarf sample}

   \subtitle{Detections and trends in the inner disk gas}

   \author{A.\,M.\,Arabhavi\,\orcidlink{0000-0001-8407-4020}\inst{1}
          \and
          I.\,Kamp\,\orcidlink{0000-0001-7455-5349}\inst{1}
          \and
          Th.\,Henning\,\orcidlink{0000-0002-1493-300X}\inst{2}
          \and
          E.\,F.\,van\,Dishoeck\,\orcidlink{0000-0001-7591-1907}\inst{3,4}
          \and
          H.\,Jang\,\orcidlink{0000-0002-6592-690X}\inst{5}
          \and
          L.\,B.\,F.\,M.\,Waters\,\orcidlink{0000-0002-5462-9387}\inst{5,6}
          \and
          V.\,Christiaens\,\orcidlink{0000-0002-0101-8814}\inst{7,8}
          \and
          D.\,Gasman\,\orcidlink{0000-0002-1257-7742}\inst{7}
          \and
          I.\,Pascucci\,\orcidlink{0000-0001-7962-1683}\inst{9}
          \and
          G.\,Perotti\,\orcidlink{0000-0002-8545-6175}\inst{2,10}
          \and
          S.\,L.\,Grant\,\orcidlink{0000-0002-4022-4899}\inst{11}
          \and
          M.\,G\"udel\,\orcidlink{0000-0001-9818-0588}\inst{12,13}
          \and
          P.-O.\,Lagage\,\orcidlink{}\inst{14}
          \and
          D.\,Barrado\,\orcidlink{0000-0002-5971-9242}\inst{15}
          \and
          A.\,Caratti\,o\,Garatti\,\orcidlink{0000-0001-8876-6614}\inst{16}
          \and
          F.\,Lahuis\,\orcidlink{}\inst{17}
          \and
          T.\,Kaeufer\,\orcidlink{0000-0001-8240-978X}\inst{18}
          \and
          J.\,Kanwar\,\orcidlink{0000-0003-0386-2178}\inst{1,19,20}
          \and
          M.\,Morales-Calder\'on\,\orcidlink{0000-0001-9526-9499}\inst{15}
          \and
          K.\,Schwarz\,\orcidlink{0000-0002-6429-9457}\inst{2}
          \and
          A.\,D.\,Sellek\,\orcidlink{0000-0003-0330-1506}\inst{3}
          \and
          B.\,Tabone\,\orcidlink{0000-0002-1103-3225}\inst{21}
          \and
          M.\,Temmink\,\orcidlink{0000-0002-7935-7445}\inst{3}
          \and
          M.\,Vlasblom\,\orcidlink{0000-0002-3135-2477}\inst{3}
          \and
          P.\,Patapis\,\orcidlink{0000-0001-8718-3732}\inst{13}
          }

   \institute{
            Kapteyn Astronomical Institute, Rijksuniversiteit Groningen, Postbus 800, 9700AV Groningen, the Netherlands\\
              \email{arabhavi@astro.rug.nl}
         \and
            Max-Planck-Institut f\"{u}r Astronomie (MPIA), K\"{o}nigstuhl 17, 69117 Heidelberg, Germany
        \and
             Leiden Observatory, Leiden University, 2300 RA Leiden, the Netherlands 
         \and
             Max-Planck Institut f\"{u}r Extraterrestrische Physik (MPE), Giessenbachstr. 1, 85748, Garching, Germany
         \and
            Department of Astrophysics/IMAPP, Radboud University, PO Box 9010, 6500 GL Nijmegen, the Netherlands
        \and
            SRON Netherlands Institute for Space Research, Niels Bohrweg 4, NL-2333 CA Leiden, the Netherlands
        \and
            Institute of Astronomy, KU Leuven, Celestijnenlaan 200D, 3001 Leuven, Belgium
        \and
            STAR Institute, Universit\'e de Li\`ege, All\'ee du Six Ao\^ut 19c, 4000 Li\`ege, Belgium
        \and 
            Department of Planetary Sciences, University of Arizona; 1629 East University Boulevard, Tucson, AZ 85721, USA.
        \and
            Niels Bohr Institute, University of Copenhagen, NBB BA2, Jagtvej 155A, 2200 Copenhagen, Denmark
        \and
            Earth and Planets Laboratory, Carnegie Institution for Science, 5241 Broad Branch Road, NW, Washington, DC 20015, USA
        \and
            Dept. of Astrophysics, University of Vienna, T\"urkenschanzstr. 17, A-1180 Vienna, Austria
        \and
            ETH Z\"urich, Institute for Particle Physics and Astrophysics, Wolfgang-Pauli-Str. 27, 8093 Z\"urich, Switzerland
        \and
            Universit\'e Paris-Saclay, Universit\'e Paris Cit\'e, CEA, CNRS, AIM, F-91191 Gif-sur-Yvette, France
        \and
            Centro de Astrobiolog\'ia (CAB), CSIC-INTA, ESAC Campus, Camino Bajo del Castillo s/n, 28692 Villanueva de la Ca\~nada, Madrid, Spain
        \and
            INAF – Osservatorio Astronomico di Capodimonte, Salita Moiariello 16, 80131 Napoli, Italy
        \and
            SRON Netherlands Institute for Space Research, PO Box 800, 9700 AV, Groningen, the Netherlands
        \and
            Department of Physics and Astronomy, University of Exeter, Exeter EX4 4QL, UK
        \and
            Space Research Institute, Austrian Academy of Sciences, Schmiedlstr. 6, A-8042, Graz, Austria
        \and
            TU Graz, Fakultät für Mathematik, Physik und Geodäsie, Petersgasse 16 8010 Graz, Austria
        \and
            Universit\'e Paris-Saclay, CNRS, Institut d’Astrophysique Spatiale, 91405, Orsay, France
             }

   \date{Received ***; accepted ***}

 
  \abstract
   {Planet-forming disks around brown dwarfs and very low-mass stars (VLMS) are on average less massive and are expected to undergo faster radial solid transport than their higher mass counterparts. \textit{Spitzer} had detected \ch{C_2H_2}, \ch{CO_2} and \ch{HCN} around these objects, but did not provide a firm detection of water. With better sensitivity and spectral resolving power than that of \textit{Spitzer}, the \textit{James Webb Space Telescope} (JWST) has recently revealed incredibly carbon-rich spectra and only one water-rich spectrum from such disks. A study of a larger sample of objects is necessary to understand how common such carbon-rich inner disk regions are and to put constraints on their evolution.}
   {We present and analyze JWST MIRI/MRS observations of 10 disks around VLMS from the MIRI GTO program. This sample is diverse, with the central object ranging in mass from 0.02 to 0.14 $M_{\odot}$. They are located in three star-forming regions and a moving group (1-10\,Myr).}
   {We identify molecular emission \hl{in all sources based on recent literature and spectral inspection,} and report detection rates. We compare the molecular flux ratios between different species and to dust emission strengths. We also compare the flux ratios with the stellar and disk properties.}
   {The spectra of these VLMS disks are extremely rich in molecular emission, and we detect the 10\,$\mu$m silicate dust emission feature in 70\% of the sample. We detect \ch{C_2H_2} and \ch{HCN} in all of the sources and find larger hydrocarbons such as \ch{C_4H_2} and \ch{C_6H_6} in nearly all sources. Among oxygen-bearing molecules, we find firm detections of \ch{CO_2}, \ch{H_2O}, and \ch{CO} in 90\%, 50\%, and 20\% of the sample, respectively. We find that the detection rates of organic molecules correlate with other organic molecules and anti-correlate with the detection rates of inorganic molecules. Hydrocarbon-rich sources show a weaker 10\,$\mu$m dust strength as well as lower disk dust mass (measured from millimeter fluxes) than the oxygen-rich sources. We find potential evidence for C/O enhancement with disk age. The observed trends are consistent with models that suggest rapid inward solid material transport and grain growth.}
   {}


   \maketitle
%
\section{Introduction}
\setlength{\tabcolsep}{4pt}
\begin{table*}[h!]
    \centering
    \caption{Summary of the source properties. The colored boxes indicate detections (green), non-detections (gray), and tentative detections (yellow) of species with \textit{Spitzer}\tablefootmark{a} indicated in the top of the last column.}
    \begin{tabular}{>{\raggedright\arraybackslash}P{30mm}>{\raggedright\arraybackslash}P{10mm}>{\raggedright\arraybackslash}P{10mm}>{\raggedright\arraybackslash}P{10mm}>{\raggedright\arraybackslash}P{8mm}>{\raggedright\arraybackslash}P{12mm}>{\raggedright\arraybackslash}P{10mm}>{\raggedright\arraybackslash}P{14mm}>{\raggedright\arraybackslash}P{5mm}>{\raggedright\arraybackslash}P{10mm}>{\raggedright\arraybackslash}P{0.05em}>{\raggedright\arraybackslash}P{0.05em}>{\raggedright\arraybackslash}P{0.05em}>{\raggedright\arraybackslash}P{0.05em}>{\raggedright\arraybackslash}P{0.05em}>{\raggedright\arraybackslash}P{0.05em}}
    \hline\hline
        Source & Short & SpTy & SFR & $d$ & log$L_{*}$ & $M_{*}$ & log$M\rm_{acc}$ & $A_{\rm v}$ & $M\rm_{dust}$ & \multirow{2}{*}{\begin{turn}{90}{\tiny $\rm C_2H_2$}\end{turn}} & \multirow{2}{*}{\begin{turn}{90}{\tiny $\rm CO_2$}\end{turn}} & \multirow{2}{*}{\begin{turn}{90}{\tiny$\rm HCN$}\end{turn}} & \multirow{2}{*}{\begin{turn}{90}{\tiny$\rm H_2O$}\end{turn}} & \multirow{2}{*}{\begin{turn}{90}{\tiny$\rm NeII$}\end{turn}} & \multirow{2}{*}{\begin{turn}{90}{\tiny $\rm H_2$}\end{turn}} \\
        2MASS- & name & & & [pc] & [$L_{\odot}$] & [$M_{\odot}$] & [$M_{\odot}$yr$\rm^{-1}$] & mag & [$M_{\oplus}$] & & & & & \\
        \hline
        J04381486+2611399 & J0438 & M7.25  & Tau   & 140.3 & -2.70 & 0.05 & -10.8$\rm ^{(1,2)}$ & 2.8 & 0.32  & \myboxx{} & \myboxx{} & \myboxx{} & \myboxxx{} & \mybox{} & \myboxxx{} \\
        J04390163+2336029 & J0439 & M6     & Tau   & 126.8 & -1.00 & 0.12 & -9.71 & 0 & 0.29    & \mybox{} & \mybox{} & \mybox{} & \myboxx{} & \myboxxx{} & \myboxx{} \\
        J11071668$-$7735532 & NC1   & M7.74  & ChaI   & 194.6  & -1.82$\rm ^{(3)}$ & 0.05$\rm ^{(3)}$ & -11.69$\rm ^{(4) }$& 0 & - & \mybox{} & \myboxx{} & \myboxx{} & \myboxx{} & \myboxx{} & \myboxx{} \\
        J11071860$-$7732516 & NC9   & M5.5   & ChaI   & 197.5 & -1.42 & 0.08 & -10.4  & 4.8 & 0.38  & \mybox{} & \myboxx{} & \myboxx{} & \myboxx{} & \myboxx{} & \myboxx{} \\
        J11074245$-$7733593 & HKCha & M5.25  & ChaI   & 191.0 & -1.25 & 0.09 & -9.62 & 2.4 & 0.89  & \mybox{} & \myboxx{} & \myboxxx{} & \myboxx{} & \myboxx{} & \myboxx{} \\
        J11082650$-$7715550 & IC147 & M5.75  & ChaI   & 195.8& -1.68 & 0.07  & -10.67 & 2.5 & <0.19 & \mybox{} & \myboxx{} & \myboxx{} & \myboxx{} & \myboxx{} & \myboxx{} \\
        J11085090$-$7625135 & Sz28  & M5.25  & ChaI   & 192.2 & -1.37 & 0.08 & -10.27 & 0.8 & <0.18 & \mybox{} & \myboxx{} & \myboxx{} & \myboxx{} & \myboxx{} & \myboxx{} \\
        J12073346$-$3932539& TWA27 & M9   & TWA & 64.4  & -2.19$\rm ^{(5)}$ & 0.02$\rm ^{(5)}$ & -11.23$\rm ^{(5)}$ & 0 & \hl{0.03}$\rm ^{(6)}$ & \mybox{} & \myboxx{} & \myboxx{} & \myboxx{} & \myboxx{} & \myboxx{} \\
        J15582981$-$2310077 & J1558 & M4.5   & UpSco & 141.1 & -1.35 & 0.14 & -9.05 & 1 & 1.19  & \mybox{} & \myboxx{} & \mybox{} & \myboxx{} & \myboxx{} & \mybox{} \\
        J16053215$-$1933159 & J1605 & M4.5   & UpSco & 152.3  & -1.54$\rm ^{(7) }$& 0.13$\rm ^{(7) }$& -9.1$\rm ^{(8) }$ & 1.6 & <0.14 & \mybox{} & \mybox{} & \mybox{} & \myboxx{} & \myboxx{} & \myboxx{} \\\hline
    \end{tabular}
    \\\justifying{The listed properties of all sources are obtained from \citet{2023ASPC..534..539M} which are updated based on the new GAIA EDR3 distances \citep{2021A&A...649A...1G}. Dust masses are based on the 0.89\,mm continuum observations (also taken from \citealt{2023ASPC..534..539M}). For sources with missing data in \citet{2023ASPC..534..539M}, the values are taken from the following references correspondingly noted in the superscripts: (1) \citet{2010ApJS..186..259R}, (2) \citet{2005ApJ...625..906M}, (3) \citet{2007ApJS..173..104L}, (4) \citet{2016A&A...585A.136M}, (5) \citet{2024AJ....167..168M}, (6) \hl{flux measurement from \citet{2017AJ....154...24R} and flux-dust mass prescription from \citealt{2023ASPC..534..539M}}, (7) \citet{2022A&A...663A..98T}, (8) \citet{2013ApJ...779..178P}.
    \tablefoottext{a}{The molecular detections with \textit{Spitzer} spectra are taken from \citet{2009ApJ...696..143P} and \citet{2013ApJ...779..178P}. For TWA27, a low resolution \textit{Spitzer} spectrum exists (PID:30540, PI: J. Houck) and was published in \citet{2008ApJ...681.1584R}, but the molecular content has not been analyzed. The detections from the reduced \textit{Spitzer} spectrum available in CASSIS \citep{2011ApJS..196....8L} have been done in this work using the criteria from \citet{2009ApJ...696..143P}.}}
    \label{tab:sources}
\end{table*}

Planets form around young stars from the dust and gas in the circumstellar disks. The physical and chemical properties of these disks vary with stellar mass and could directly affect the planet formation process. \hl{Infrared observations have revealed that disks surrounding the lowest-mass stars exhibit peculiar, carbon-rich gas compositions that differ greatly from those around higher-mass stars} \citep{2009ApJ...696..143P,2013ApJ...779..178P,2023NatAs...7..805T,2024Sci...384.1086A}. These studies focus on disks around central objects with masses up to 0.15 solar masses, which we refer to as very low-mass stars or `VLMS'. Planet population synthesis models and observations indicate that rocky planets have the highest occurrence rates around objects with masses much lower than those of Sun-like stars \citep{2021A&A...656A..72B,2021A&A...653A.114S}. This makes the study of VLMS disks very interesting for understanding the formation of rocky planets. The composition of the inner regions of these disks can provide strong constraints on the composition of the planets that form around such stars.

Disks around VLMS differ from those around higher mass T\,Tauri stars in several ways. Due to the low disk masses and the stellar luminosities, the ice lines in the VLMS disks are quite close-in (\citealt{2016ApJ...831..125P}, \citealt{2017A&A...601A..44G}). For example, models show that the water and CO icelines around a typical T\,Tauri disk are at about 1\,au and 20\,au respectively \citep{2017A&A...607A..41K}, while the same icelines around a typical VLMS are at about 0.1\,au and 2\,au \citep{2017A&A...601A..44G}. Due to such small distances between icelines and the very short dynamic timescales in the VLMS disks \citep{2013A&A...554A..95P,2023arXiv231009077V}, the radial transport processes have a more significant impact on their disk composition compared to T\,Tauri disks \citep{2023A&A...677L...7M}. Observations such as those from the \textit{Hubble Space Telescope} \citep{2012ApJ...755..154M} and the \textit{Very Large Telescope} \citep{2010A&A...510A..72F} show that the accretion rates in VLMS disk systems fall off quickly by several orders of magnitude within a few Myr, whereas those in the T\,Tauri disks decrease much slower. Observations also show that the disks around VLMS have, in general, weaker silicate dust features compared to the higher mass counterparts, indicating a more efficient grain growth \citep{2005Sci...310..834A,2009ApJ...696..143P}. Further, the properties of the central objects also differ significantly in terms of their stellar activity, X-ray luminosity, temperature, and stellar spectral energy distribution (\citealt{2007A&A...468..353G}, \citealt{2000ApJ...538..363B}, \citealt{2003ApJ...582.1109W}, \citealt{2012ARA&A..50...65L}, \citealt{2010ApJS..186...63R}).

\textit{Spitzer} Infrared Spectrograph (\textit{Spitzer}-IRS) observations \citep{2009ApJ...696..143P,2013ApJ...779..178P} of the inner regions of the VLMS disks \citep[$\rm \lesssim0.1\,au$ from the central object,][]{2007ApJ...659..680K} have shown an underabundance of HCN relative to $\rm C_2H_2$ when compared to the inner disks of the higher mass counterparts ($\rm \lesssim1\,au$ from the central object) and only tentative detections of water in two young disks in the Taurus star-forming regions (SFRs). Observations with the higher resolution and sensitivity of the Mid-InfraRed Instrument (MIRI) on the \textit{James Webb Space Telescope} (JWST) have enabled detailed characterization of the VLMS, revealing a rich hydrocarbon chemistry including isotopologues of some species \citep[][Morales-Calder\'on et al. subm.]{2023NatAs...7..805T,2024Sci...384.1086A,2024A&A...689A.231K}. These objects generally show strong hydrocarbon emission and sometimes even hydrocarbon molecular pseudo-continua, but do not show firm detections of water or OH, indicating the C/O ratio is larger than unity in the emitting regions \citep{2023NatAs...7..805T,2024Sci...384.1086A}. \citet{2023ApJ...959L..25X} reported a water-rich disk around the source Sz\,114 ($M_*=0.17\,M_{\odot}$) and suggest that this source could be at an early evolutionary stage compared to the other VLMS disks, while \citet{2024arXiv241205535L} presented a very carbon-rich MIRI spectrum of a 30\,Myr old VLMS disk. \citet{2025arXiv250411424P} present a MIRI spectrum of a highly inclined brown dwarf disk which appears to be water-rich. \hll{More recently, \citet{2025arXiv250513714F} reported a hydrocarbon-rich disk around a planetary-mass object ($<$0.01\,$M_{\odot}$) with a MIRI spectrum remarkably similar to the VLMS disk presented by \citet{2024Sci...384.1086A}.}

Two scenarios are proposed to explain the hydrocarbon-rich inner disk \citep{2023NatAs...7..805T,2023A&A...677L...7M,2024Sci...384.1086A}: i) carbon enrichment through carbon grain destruction, ii) oxygen depletion by disk transport processes or dust traps. In the former, the carbon trapped in the dust is released to the gas phase by sublimation, combustion, chemosputtering, or photoprocesses, leading to a carbon-rich gas in the inner disk. In the latter, efficient transport processes in these VLMS disks \citep{2013A&A...554A..95P} can lead to early preferential accretion of the oxygen-bearing molecules onto the central object fed by the oxygen-dominated ices in the midplane beyond the snowline \citep{2023A&A...677L...7M}. This would lead to an oxygen-depleted gas in the inner disk over time. In both of these scenarios, the enhanced C/O gas composition (C/O>1) leads to efficient formation of hydrocarbons either in the gas phase \citep{2024A&A...689A.231K} or on grain surfaces (provided suitable dust temperatures) (\citealt{2007ApJ...655L..49W}, \citealt{2019NatAs...3..568H}). 

Inner disk substructures can influence the local chemical compositions and play an important role in explaining the observed infrared spectra of such sources. \citet{2023ApJ...954...66K}, \citet{2024A&A...686L..17M}, \citet{2024arXiv240209342L}, and \citet{2024arXiv241201895S} show that the nature of the gaps, particularly whether the gaps are due to pebble traps, planets or photoevaporative winds, can strongly influence the inner disk elemental composition. Kanwar et al. (subm.) carry out thermochemical modeling introducing a gap that splits the disk into a high C/O inner disk and a more normal outer disk and can reproduce the observed JWST carbon-rich spectrum including pseudo continuum and \ch{CO_2} emission in the absence of strong water emission. 

Due to the low disk masses, luminosities, and small spatial scales, there is a lack of spatially resolved observations of the inner disks of these systems. Deeper and more extensive studies, such as with the Atacama Large Millimeter/submillimeter Array (ALMA), should be used to provide more detailed information for the brighter VLMS disks. 

In this paper, we present and analyze the JWST MIRI spectra of disks around 10 VLMS, focusing on the gas detections. \citet{2025ApJ...984L..62A} provide a detailed discussion on detections of water in the sample. We present the sample, the observations, and the data reduction process in Sect.\,\ref{sec:observations}. We discuss the molecular detections and the detection rates in Sect.\,\ref{sec:detections}. In Sect.\,\ref{sec:trends2} and \ref{sec:spectralappearance}, we present trends observed in the molecular flux ratios and notes on individual sources, respectively. The implications of our findings are discussed in Sect.\,\ref{sec:discussion} and finally the conclusions are presented in Sect.\,\ref{sec:conclusions}.

\section{The sample, observations and data reduction}
\label{sec:observations}
\subsection{The sample}
The observations are part of \textbf{M}IRI mid\textbf{IN}frared \textbf{D}isk \textbf{S}urvey (MINDS) program, a JWST Guaranteed Time Observations (GTO) program (ID: 1282, PI: Th. Henning, \citealt{2024PASP..136e4302H}, \citealt{2023FaDi..245..112K}). This GTO program includes 10 VLMS sources, of which nine have been observed, and are presented in this work \citep[see also][Morales-Calder\'on et al. subm]{2023NatAs...7..805T,2024Sci...384.1086A,2024A&A...689A.231K,2025arXiv250411424P}. We also include a brown dwarf disk, TWA27A (further referred to as TWA27), from the exoplanet GTO program (PID:1270, PI:S.Birkmann, also see Patapis et al. subm.). The properties of the sample are summarized in Table\,\ref{tab:sources}. The spectral type of the sources ranges from M4.5 to M9. These sources are located in three star-forming regions - Taurus, Chamaeleon, and Upper Sco, and one moving group, the TW Hydrae Association. Some of the sources have high visual extinction values, e.g. $A_{\rm v}$=4.8\,mag for NC9 (see Table\,\ref{tab:sources}). One of the sources, J0438, is a highly inclined disk ($\sim$70$\rm^o$, \citealt{2006ApJ...645.1498S,2007ApJ...666.1219L}). All of the sources in our sample have been previously observed with \textit{Spitzer}-IRS \citep{2008ApJ...681.1584R,2009ApJ...696..143P,2013ApJ...779..178P}. The spectra showed both dust features and gas emission lines. Table\,\ref{tab:sources} summarizes the molecular and atomic emissions detected with \textit{Spitzer}. All sources have been observed with ALMA in the continuum at 0.89\,mm \citep{2016ApJ...831..125P,2016ApJ...827..142B,2017AJ....154...24R,2018AJ....155...54W}, except NC1 for which the observation had failed. In these observations, IC147, Sz28, and J1605 have not been detected. The estimated dust masses (or the upper limits) are listed Table\,\ref{tab:sources}. Only J0438, J1558, and TWA27 have been firmly detected in millimeter gas emission in the $^{12}$CO\,$J=3-2$ line with ALMA (\citealt{2016ApJ...827..142B}, \citealt{2017AJ....154...24R}, \citealt{2025arXiv250411424P}). 1.3\,mm continuum observations have been performed with IRAM 30\,m single-dish telescope and the Submillimeter Array (SMA) telescope for the two Taurus sources and with ALMA for J0438 \citep{2013ApJ...771..129A,2006ApJ...645.1498S,2018AJ....155...54W}. Compared to their higher-mass counterparts, the VLMS disks are smaller and fainter, which limits their observability. 

\subsection{Observations}
The sources were observed with the Mid-InfraRed Instrument \cite[MIRI;][]{miri_rieke2015PASP,Wright2015,Wright2023} of JWST with the Medium Resolution Spectroscopy \cite[MRS;][]{Wells2015,Argyriou2023} mode. This involves four Integral Field Units (IFUs): channel 1 (4.9$-$7.65~$\mu$m), channel 2 (7.51$-$11.71~$\mu$m), channel 3 (11.55$-$18.02~$\mu$m), and channel 4 (17.71$-$27.9~$\mu$m), and each channel is composed of three sub-bands: SHORT (A), MEDIUM (B), and LONG (C), leading to a total of twelve wavelength bands. The sources were observed with target acquisition and were observed in FASTR1 readout mode with a 4-point dither pattern in the positive direction, for a total exposure time per sub-band per source of $\sim$1232\,s. TWA27, however, was observed with the extended source 4-point dither pattern in the negative direction for a total exposure time per sub-band of $\sim$844\,s.

\subsection{Data reduction}
\label{sec:datareduction}
We used a hybrid pipeline \citep[v1.0.3,][]{2024ascl.soft03007C} relying on the standard JWST pipeline \citep[v1.14.1][]{Bushouse2023} using CRDS context (\texttt{jwst\_1254.pmap} file), and complemented with routines from the VIP package \citep{GomezGonzalez2017,Christiaens2023}, to reduce the data. The pipeline is structured around three main stages that are the same as in the JWST pipeline, namely \texttt{Detector1}, \texttt{Spec2} and \texttt{Spec3}. After the first stage (\texttt{Detector1}), stray light is corrected, and a background estimate subtracted. To remove the background, we carried out a direct pair-wise dither subtraction. This method is more suited to fainter sources (which is the case for all our VLMS sources) where the resulting PSF overlap is minimal and reduces the noise level in the spectrum, but can lead to a minor flux discrepancy with prior \textit{Spitzer} measurements due to self-subtraction (this minor discrepancy will not affect our analysis in this work). Since TWA27 was observed in an extended source dither pattern, the half-integral dither offset leads to self-subtraction while using the direct dither subtraction to remove the background. This results in a drop in flux in the final spectrum. We compared this with the spectrum obtained by measuring the background in an annulus around the source. The latter is noisier, but matches well with the flux levels reported with \textit{Spitzer}. We used the flux levels from the annulus to rescale the less noisier spectrum obtained from direct dither subtraction. This does not affect the relative or absolute molecular and dust flux levels. \texttt{Spec2} was then used with default parameters, but the stray light correction and background subtraction were skipped.

The outlier detection step in \texttt{Spec3} was skipped, and replaced by a custom VIP-based bad pixel correction routine applied before \texttt{Spec3}. The bad pixels are identified through sigma-filtering, and corrected with a Gaussian kernel. \hl{This significantly reduced the number of spikes otherwise present in the spectra extracted with early versions of the official JWST pipeline\footnote{Tests carried out after completing the analysis presented in this work revealed that as of v1.14 onward, the bad pixel correction performed using default parameters of the official pipeline now leads to similar quality spectra as obtained with VIP-based bad pixel correction.}.} VIP-based routines are then also used for the identification of the star centroid in the spectral cubes produced at stage 3. The centroid location is identified with a 2D Gaussian fit in a weighted mean image for each band's spectral cube, where the weights are set to be proportional to each frame's integrated flux in the central part of the field. The identified centroid locations are subsequently used for aperture photometry: the spectrum is extracted by summing the signal in a $2$$\times$FWHM aperture centered on the source, where the FWHM is equal to 1.22 $\lambda / D$, with $\lambda$ the wavelength and $D \sim 6.5$m the diameter of the telescope. Aperture correction factors are applied to account for the flux loss as presented in \citet{Argyriou2023}.

For the analysis of the spectra, we require molecular fluxes, i.e., line fluxes above the dust continuum. For this, we determine the continua using the procedure described in \citet{Temmink2024} which uses the \texttt{pybaselines} package \citep{erb_2022_7255880}. This allows for a reproducible continuum definition instead of tracing the dust emission by eye. The method first estimates a continuum level using Savitzky-Golay filter and masks strong line emission above 2$\sigma$. Further, \citet{Temmink2024} mask the 3$\sigma$ downward spikes. The VLMS spectra are generally line-rich, and can show molecular pseudo-continuum emission. This leads to difficulties in 1) distinguishing between broad dust continuum features and molecular continuum, 2) identifying the dust continuum level to mask the downward spikes. So we skip the masking of the downward spikes. We define two continua for each source, corresponding to broad and narrow widths of the Savitzky-Golay filter. Subtracting the continuum defined with the narrow width subtracts the dust (including the optically thin dust features) as well as the molecular pseudo-continuum, while the broad width continuum retains the broad features. The latter is used to measure the dust strengths and the molecular fluxes in case of molecular pseudo-continuum. More details on the continuum definition method can be found in Sect.\,2.2 of \citet{Temmink2024}.

The final reduced spectra, and comparison with the \textit{Spitzer} spectra are presented in App.\,\ref{sec:continuum} and \ref{sec:spitzer}.

\begin{figure*}[!ht]
    \centering
    \includegraphics[width=\linewidth]{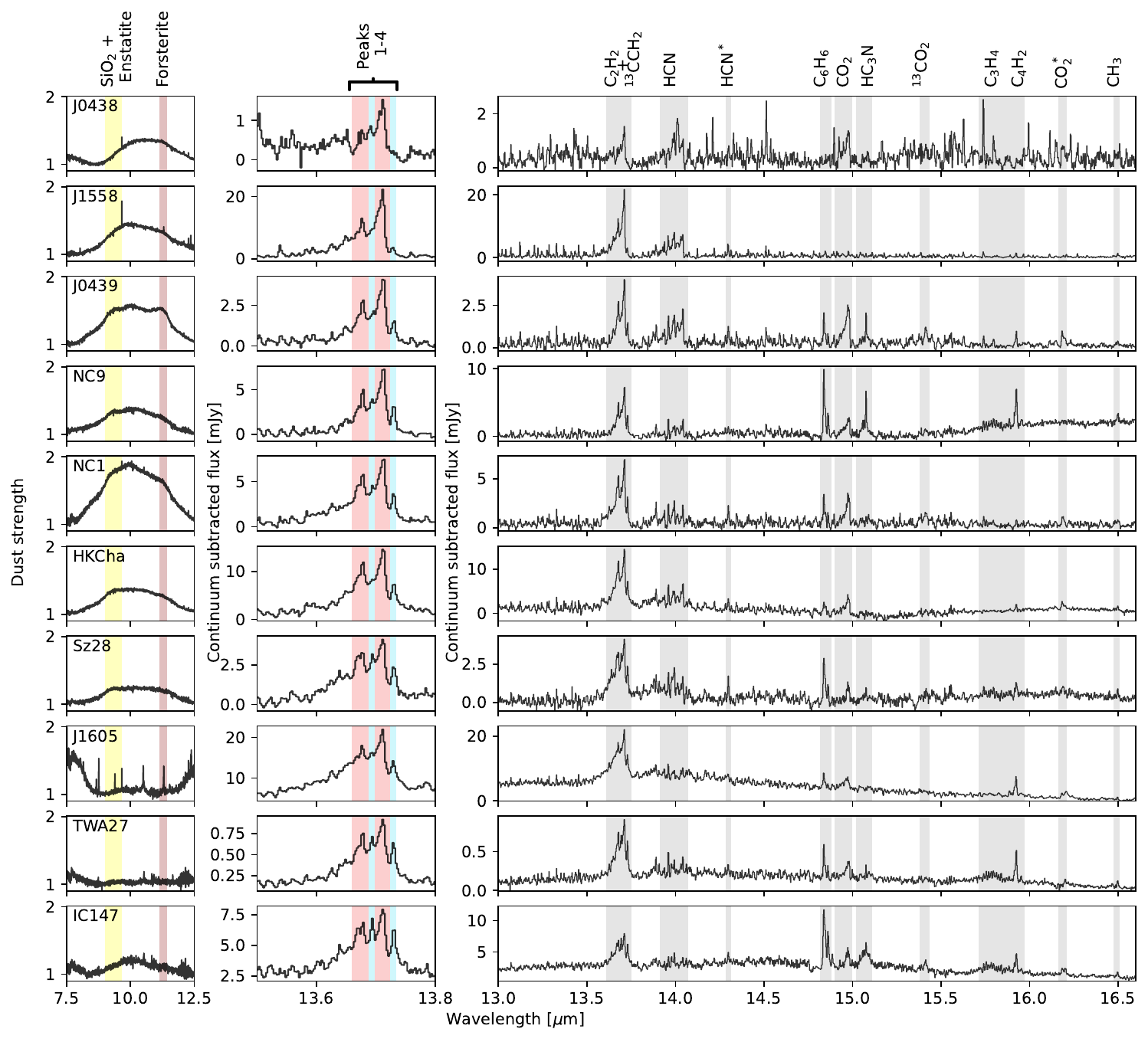}
    \caption{Summary of dust and gas observations. The left panels show continuum-normalized dust strengths highlighting \ch{SiO_2}, enstatite, and forsterite emission features. The middle panels show a zoom in on the observed \ch{C_2H_2} features highlighting four peaks corresponding to the main and rare isotopologues in red and blue respectively. The right panels show the continuum-subtracted hydrocarbon-rich wavelength region of the spectra. The grey regions highlight the main $Q$-branches of molecules in this wavelength region. \ch{HCN^*} and \ch{CO_2^*} indicate hot bands $\nu_2$:1-0 and $\nu_1\nu_2\nu_3$:100-010, respectively. The spectra are ordered by increasing flux ratios of the \ch{^{13}CCH_2} and \ch{C_2H_2} peaks (see Sect.\,\ref{sec:trends2}) and are normalised to a distance of 150\,pc.}
    \label{fig:features_summary}
\end{figure*}

\subsection{Extinction}
The spectra are affected by extinction along the line of sight. The extinction estimates based on ultraviolet (UV), visible and near infrared spectroscopy presented by \citet{2008ApJ...681..594H},  \citet{2014ApJ...786...97H}, \citet{2016A&A...585A.136M}, and \citet{2017A&A...604A.127M} are listed in Table\,\ref{tab:sources}. J0438 is a highly inclined disk whose extinction estimate is expected to be largely influenced by its high inclination ($>$70$\rm^o$, \citealt{2006ApJ...645.1498S}), but also by any foreground extinction. Hence, the reported value ($A_v$=2.8\,mag) for J0438 should be taken as an upper limit. 

For the rest of the paper, we de-redden the spectra using the extinction law from \citet{2023ApJ...950...86G}. A comparison of the spectra before and after de-reddening is presented in App.\,\ref{sec:continuum}. Since most of the molecular emission analyzed in this paper occurs in a narrow wavelength range, well separated from the silicate features, the accuracy of the extinction values and the extinction curves do not affect our conclusions.

\begin{figure*}[h!]
    \centering
    \includegraphics[width=\linewidth]{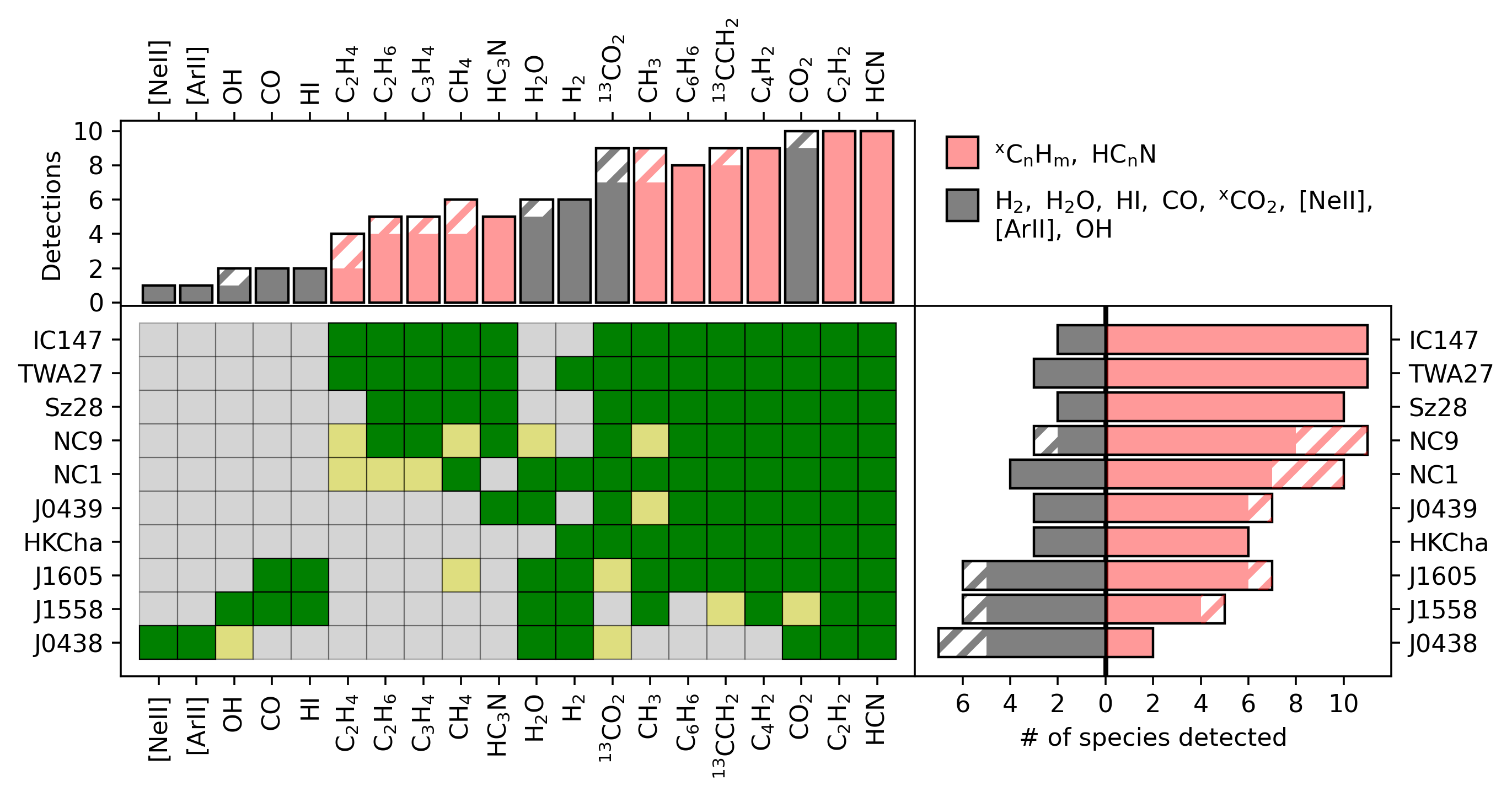}
    \caption{Summary of dust and gas detections in the sample. Green square refers to a firm detection, yellow indicates a tentative detection and light gray square indicates a non-detection. The top panel shows the number of detections for each species. The right panel shows the number of species of two groups detected in each source. The inorganic species (dark gray) includes \ch{H_2}, \ch{H_2O}, \ch{HI}, \ch{CO}, \ch{CO_2}, \ch{^{13}CO_2}, [Ne\,$\textsc{II}$], and OH. The organic species (faded red) includes all the hydrocarbons, and cyanomolecules. The hatched region in the top and right panels show tentative detections.}
    \label{fig:detectionrates}
\end{figure*}

\section{Detections and detection rates}
\label{sec:detections}

The MIRI spectra of our sample show several molecular emission bands along with dust features of varying strengths. The detection criteria and detection of individual species in each source are provided in App.\,\ref{app:detections}. Here we summarize the detections and the detection rates.

\subsection{Detection rates}
Figure.\,\ref{fig:features_summary} provides an overview of the dust and gas features. The left panels show the continuum-normalized dust strengths of the optically thin emission. Seven sources (J0438, J0439, J1558, NC9, NC1, HKCha, and Sz28) show prominent silicate emission (Fig.\,\ref{fig:features_summary}), while the remaining three sources (J1605, TWA27, IC147) show almost flat 10\,$\mu$m regions. IC147 shows a shallow feature with a peak at $\sim$10\,$\mu$m whose shape is quite distinct from the rest of the detected dust features. Moreover \citet{2024Sci...384.1086A} show that emission from a large column density of \ch{C_2H_4} can reproduce the broad shape before de-reddening. 

In general, the observed 10\,$\mu$m dust shapes are flat-topped and weaker than those observed in the T\,Tauri disks (also see Grant et al. in prep.), in-line with previous \textit{Spitzer} observations (\citealt{2005Sci...310..834A}, \citealt{2007ApJ...659..680K}). We find peaks at $\sim$9.2\,$\mu$m corresponding to \ch{SiO_2} and/or enstatite and 11.3\,$\mu$m corresponding to forsterite. A more detailed analysis of these dust features will be presented in Jang et al. (in prep). In this paper, we focus on the gas composition of these inner disks.

All of the sources are rich in gas species, as shown in the middle and right panels in Fig.\,\ref{fig:features_summary}. \ch{C_2H_2} and \ch{HCN} are ubiquitously detected across all the sources due to the better sensitivity of JWST, compared to detection rates of 41\,\% and 22\,\%, respectively, deduced from \textit{Spitzer} spectra \citep{2009ApJ...696..143P,2013ApJ...779..178P}. \ch{CO_2} is detected in all sources with the exception of J1558, in which the detection is tentative (Fig.\,\ref{fig:remaindetect}). Considering the entire sample, we detect \ch{H \textsc{i}}, [Ar\,$\textsc{ii}$], [Ne\,$\textsc{ii}$], \ch{H_2}, \ch{OH}, \ch{H_2O}, \ch{CO}, \ch{CO_2}, \ch{HCN}, \ch{HC_3N}, \ch{CH_3}, \ch{CH_4}, \ch{C_2H_2}, \ch{C_2H_4}, \ch{C_2H_6}, \ch{C_3H_4}, \ch{C_4H_2}, and \ch{C_6H_6}, along with two isotopologues - \ch{^{13}CO_2} and \ch{^{13}CCH_2}. Consistent with the \textit{Spitzer} findings \citep[e.g.][]{2009ApJ...696..143P}, the \ch{HCN} fluxes are typically weaker than the \ch{C_2H_2} fluxes. The VLMS sample studied by \citet{2013ApJ...779..178P} with \textit{Spitzer} showed a [Ne\,$\textsc{ii}$] detection rate of 2/8, while it is only 1/10 in our sample. \citet{2013ApJ...779..178P} reported a tentative detection of [Ne\,$\textsc{ii}$] in J0439, but we do not detect [Ne\,$\textsc{ii}$] in the MIRI spectrum.

Figure \ref{fig:detectionrates} summarizes the detections and detection rates of different species. The lower left panel shows the emission features detected in each source. The top panel shows the detection rates of organic molecules in light red and inorganic molecules in gray. In general, these spectra are rich in organic molecules. If we include the tentative detections, more than 50\% of the detected molecules are purely hydrocarbons, and more than 80\% contain carbon. While we detect the hydrocarbons such as \ch{C_2H_2}, \ch{C_4H_2}, and \ch{C_6H_6} in 10, 9, and 8 sources in the sample of 10 objects, we detect oxygen-bearing molecules such as \ch{CO_2}, \ch{H_2O}, and \ch{CO} in 9, 5, and 2 sources respectively (see \citealt{2025ApJ...984L..62A} for a more detailed discussion on detections of water in the sample). \ch{H_2}, and \ch{H_2O} are the most commonly detected non-carbon-bearing molecules. \ch{C_2H_4} is the least commonly detected organic molecule (only TWA27, and IC147). While three sources (J0438, J1558, J1605) show \ch{H\textsc{i}}, \ch{H_2} and \ch{H_2O} emission simultaneously, \ch{CO} is detected only in the two UpperSco objects, J1605 and J1558. Except for J1605, J1558, and NC1, all the other sources are dominated by the stellar photospheric absorption of \ch{CO} and \ch{H_2O} at shorter wavelengths ($\rm \lesssim7\ \mu m$). Across the sample, benzene (\ch{C_6H_6}) is the only aromatic hydrocarbon detected, and we do not find any polycyclic aromatic hydrocarbons (PAHs) in these spectra.

\subsection{Notes on spectral appearance}
\label{sec:spectralappearance}
\textit{J0438}: It is a highly inclined disk with a peculiar 10\,$\mu$m shape showing dust in both absorption and emission. The MIRI spectrum appears water-rich closer to T\,Tauri spectra in appearance (water was tentatively detected in its \textit{Spitzer} spectrum, \citealt{2013ApJ...779..178P}). \ch{C_2H_2} is the only hydrocarbon that is detected and the \ch{CO_2} flux is also weaker than many water lines. This is the only source where [Ne\,$\textsc{ii}$] and [Ar\,$\textsc{ii}$] are detected, (see \citealt{2025arXiv250411424P} for a full discussion on this source).\\
\textit{J1558}: It is a \ch{C_2H_2}-bright source. The continuum-subtracted peak flux of \ch{C_2H_2} is $\sim$22.2\,mJy, which is almost identical to the peak flux of \ch{C_2H_2} in the J1605 spectrum (both are UpperSco sources, see Fig.\,\ref{fig:features_summary}). However, the \ch{^{13}CCH_2} feature is absent or very weak, implying that \ch{C_2H_2} in J1558 has column densities of at least an order of magnitude lower than what is reported for J1605 \citep{2023NatAs...7..805T}. Such bright emission at low column densities requires a large emitting area or high temperatures. The J1558 spectra is also water-rich, but surprisingly \ch{CO_2} is absent or very weak. This is opposite to what is observed in the rest of the sample, where \ch{CO_2} is more readily detectable than water.\\
\textit{J0439}: This is the only source where we clearly detect several organic molecules as well as rotational water lines. \\
\begin{figure}[!h]
    \centering
    \includegraphics[width=0.8\linewidth]{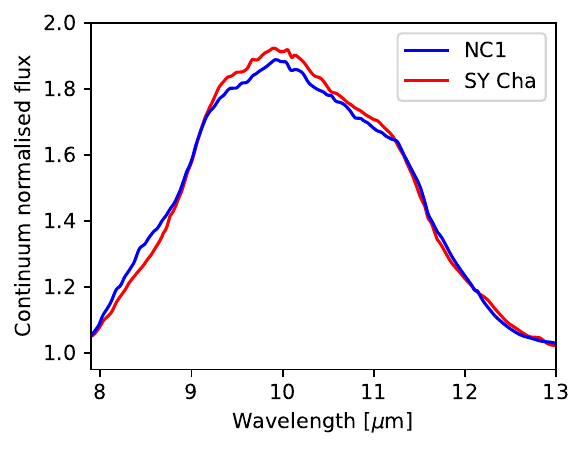}
    \caption{Comparison of the 10\,$\mu$m dust feature of NC1 (VLMS) and SY Cha (T\,Tauri). Here the continuum normalised flux is shown. The observed MIRI spectrum of SY Cha is taken from \citet{2024ApJ...962....8S}.}
    \label{fig:NC1SYC}
\end{figure}
\textit{NC1}: This source has the strongest 10\,$\mu$m dust feature. The dust feature shape resembles closely some of the T\,Tauri disks. Figure\,\ref{fig:NC1SYC} compares the dust feature with SY\,Cha \citep{2024ApJ...962....8S}. While this source shows infrared excess and ro-vibrational water emission shortward of 8\,$\mu$m, \ch{CO} emission is not detected. This source also shows emission from several organic molecules (see Morales-Calder\'on et al. subm. for more details).\\
\textit{NC9, HKCha, and Sz28}: These sources have similar dust shapes and strengths but differ in terms of their molecular detections. \\
\textit{J1605}: This source has one of the brightest \ch{C_2H_2} emission strengths in a VLMS disk and \ch{C_2H_2} is extremely optically thick \citep{2023NatAs...7..805T}. It also has the flattest 10\,$\mu$m region of all our spectra without an optically thin dust feature. The number of organic molecules detected in J1605 is smaller than in many other sources in our sample.\\
\textit{TWA27 and IC147}: These two sources show the largest number of organic molecules in emission. The overall shape of the MIRI spectra are also similar. Both sources show emission from \ch{C_2H_4}, which coincides with the 10\,$\mu$m dust feature. It is not fully clear whether the broad shapes at these wavelengths are entirely \ch{C_2H_4} or partly due to dust.

\begin{figure}[!ht]
    \centering
    \includegraphics[width=0.65\linewidth]{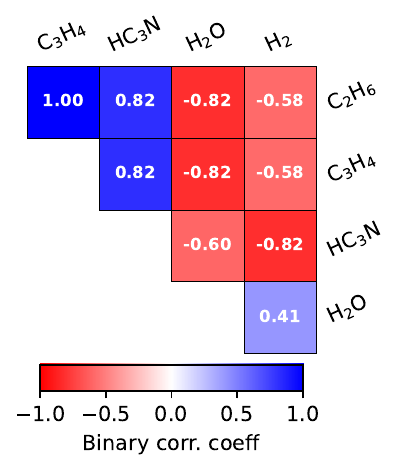}
    \caption{Binary correlation coefficients of a few species that show strong (anti-)correlations. Correlation coefficients (and the p-values) for a more extended list of species are shown in Fig.\,\ref{fig:correlations}.}
    \label{fig:corr_zoom}
\end{figure}

\subsection{Correlations and anti-correlations in detection rates}
The right panel of Fig.\,\ref{fig:detectionrates} shows the number of organic and inorganic molecules detected in each source. The number of organic molecules detected in most sources is much larger than that of inorganic molecules. Exceptions of this are the highly-inclined disk J0438 and the \ch{C_2H_2}-bright disk J1558.

A quantitative measure of the trends in the gas detections can be obtained by calculating correlation coefficients. The right panel of Fig.\,\ref{fig:detectionrates} shows that the number of organic and inorganic molecules detected in each source are anti-correlated, with a Pearson correlation coefficient of \hl{-0.78} (p-value \hl{0.007}). Figure\,\ref{fig:correlations} shows the binary correlation coefficients $\mathrm{\phi}$ \citep{Yule1912} and the p-values between the species detected in the sample. A strong correlation between a molecule pair implies that it is more probable for both molecules to be either detections or non-detections. On the other hand, an anti-correlation between a molecule pair implies that it is more probable that only one of the molecules is detected at a time. Less dominant (relative to \ch{C_2H_2}) organic molecules such as \ch{C_3H_4}, \ch{C_2H_6}, and \ch{HC_3N} are strongly correlated (see Fig.\,\ref{fig:corr_zoom}). Inorganic molecules such as \ch{H_2} and \ch{H_2O} are anti-correlated with the organic molecules. Since we detect \ch{CO_2} in all sources, it is not surprising to find that also \ch{^{13}CO_2} correlates with organic molecules (Fig.\,\ref{fig:correlations}). 

\section{Trends in the sample}
\label{sec:trends2}

\subsection{\ch{^{13}CCH_2} to \ch{C_2H_2} flux ratio as a measure of column density}

Estimating the column densities, temperatures, and emitting radii of the molecules using 0D slab models\footnote{A single point model with a column density, temperature, and an emitting radius.}, as done with \textit{Spitzer} observations, would be beneficial for a quantitative analysis. However, several molecular features observed in the spectra cannot be reproduced by models due to lack of molecular data (e.g. \citealt{2023NatAs...7..805T}, \citealt{2024Sci...384.1086A}, \citealt{2024A&A...689A.231K}). In addition, studies of JWST observations of disks around VLMS such as \citet{2023NatAs...7..805T} and \citet{2024Sci...384.1086A} have shown that the fitting procedure is not trivial for these rich spectra due to spectral overlap of multiple molecules. Moreover, a single 0D model might not be able to reproduce the observed molecular features. Analysis using more sophisticated tools such as 1D retrieval tools (e.g. CLIcK - \citealt{2019A&A...623A.106L}, DuCKLinG - \citealt{2024arXiv240506486K,2024A&A...690A.100K}) is left to a future work. Consequently, we limit ourselves to the analysis using the integrated line fluxes in this work. Further, the spectra are rich in molecules, and the molecules individually emit in a very large wavelength range, which makes estimating the total flux of a molecule impossible without performing slab model fits. Therefore, we rely only on the fluxes of the $Q$-branches for the rest of the analysis.

The middle panels in Fig.\,\ref{fig:features_summary} show a zoom-in on the \ch{C_2H_2} features of the sample. We identify four peaks: 
\begin{enumerate}
    \item 13.679\,$\mu$m \ch{C_2H_2} - $v_4+v_5:1,1-1,0$
    \item 13.691\,$\mu$m \ch{^{13}CCH_2} - $v_4+v_5:1,1-1,0$
    \item 13.712\,$\mu$m \ch{C_2H_2} - $v_5:1-0$
    \item 13.730\,$\mu$m \ch{^{13}CCH_2} - $v_5:1-0$
\end{enumerate}
The relative strengths of these peaks exhibit considerable diversity across the sample. The width at the base also varies. Each peak mentioned above is a band head with a tail towards the shorter wavelengths, which results in flux contributions of each band to the remaining peaks at the shorter wavelengths. Peaks 3 and 4 are the least contaminated peaks corresponding to the primary and secondary isotopologue of \ch{C_2H_2} whose integrated fluxes (between 13.698 - 13.723\,$\mu$m, and 13.723 - 13.734\,$\mu$m respectively) will be referred to as $F_{\rm C_2H_2}$ and $F_{\rm ^{13}CCH_2}$ in the rest of the paper. The sources in Fig.\,\ref{fig:features_summary} are ordered according to the ratio of these fluxes ($F_{\rm ^{13}CCH_2}$/$F_{\rm C_2H_2}$).

As the gas column density increases, the optical depth also increases. Consequently, at high gas column densities, the integrated flux $F_{\rm C_2H_2}$ saturates. \hl{Assuming that the isotopologue \ch{^{13}CCH_2} emits from the same region as \ch{C_2H_2}, t}he abundance of the isotopologue \ch{^{13}CCH_2} is expected to be a factor 1/35 lower than \ch{C_2H_2} \citep[assuming the interstellar \ch{^{12}C}/\ch{^{13}C} ratio of $\sim$70,][]{2009ApJ...693.1360W}, which makes it optically thinner. Therefore, the peak associated with that, $F_{\rm ^{13}CCH_2}$, is less saturated at those high column densities. \hl{If the \ch{C_2H_2} and \ch{^{13}CCH_2} gas columns extend below the optically thick $\tau_{\rm C_2H_2}$=1 surface, $F_{\rm ^{13}CCH_2}$} continues to increase with the increasing column density before saturation. Assuming that the isotopic fractionation of carbon is uniform across the disk, the $F_{\rm ^{13}CCH_2}/F_{\rm C_2H_2}$ ratio would be least affected by the radial extent of the hydrocarbon emission. However, the flux ratio would strongly reflect the vertical gas composition above the optically thick dust layer (the depth of this dust layer would be determined by the dust size distribution and dust settling). We can leverage the ratio of the integrated fluxes of these two peaks as a measure of the column density or the gas optical depth probed by MIRI. Since we expect to probe large column densities of \ch{C_2H_2} and \ch{^{13}CCH_2}, we also expect a large contribution of higher excited bands to the $Q$-branch peaks. Due to the unavailability of spectroscopic data of \ch{^{13}CCH_2} beyond the fundamental band, a quantitative analysis of the observed fluxes to determine column densities is not feasible. Qualitatively, a higher value of $F_{\rm ^{13}CCH_2}$/$F_{\rm C_2H_2}$ indicates a larger column density of gas being probed. 

\hl{While we assume uniform isotopic fractionation of carbon in case of \ch{C_2H_2} and \ch{^{13}CCH_2}, preferential shielding from ultraviolet radiation, or chemical fractionation could lead to deviations in the local \ch{^{12}C}/\ch{^{13}C} ratio. In such cases, the range of $F_{\rm ^{13}CCH_2}/F_{\rm C_2H_2}$ ratios and the related trends (Sect.\,\ref{sec:miritrends} \& \ref{sec:stellardisktrends}) reported for our sources could be partly due to these fractionation processes.}

\begin{figure}[h!]
    \centering
    \includegraphics[width=0.65\linewidth]{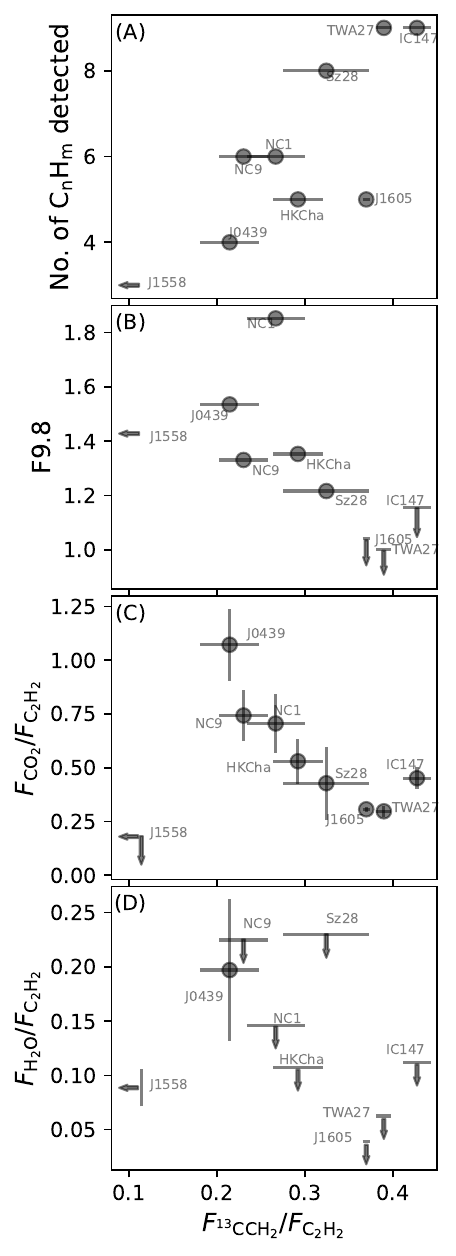}
    \caption{Trends in the MIRI observations. Panels A-D show the number of hydrocarbons (\ch{C_nH_m}) detected, integrated dust strength, the \ch{CO_2} to \ch{C_2H_2} flux ratios, and the \ch{H_2O} to \ch{C_2H_2} flux ratios against the flux ratio $F_{\rm ^{13}CCH_2}$/$F_{\rm C_2H_2}$, along with the errorbars. The molecules with tentative or non-detections are indicated by the upper limits. In case of non-detections, we report the 1$\sigma$ upper limits. All errorbars are 1$\sigma$ with an assumption of a signal-to-noise ratio of 100.}
    \label{fig:C2H2trends}
\end{figure}

\subsection{Trends in the MIRI observations}
\label{sec:miritrends}
In Fig.\,\ref{fig:C2H2trends}, we investigate the detection rate of organic molecules, the dust strengths and the molecular flux ratios as a function of $F_{\rm ^{13}CCH_2}$/$F_{\rm C_2H_2}$. \hl{Using the JWST Exposure Time Calculator\footnote{\url{https://jwst.etc.stsci.edu/}} and adopting the same observational setup—including the MIRI exposure times—as used for our observations, we estimate signal-to-noise ratios (SNRs) ranging from 100 to 270 in the wavelength regions where the line fluxes are measured, based on the observed flux levels. To be conservative, we choose the lower limit of the SNR range (=100) for the errorbars and upper limits.} More details on the SNR estimation is discussed in Jang et al. (in prep.). We do not include J0438 because it is highly inclined, and the $F_{\rm ^{13}CCH_2}$/$F_{\rm C_2H_2}$ ratio would have large errorbars.

\subsubsection{Number of hydrocarbon species detected}
Except for J1605, a higher $F_{\rm ^{13}CCH_2}$/$F_{\rm C_2H_2}$ ratio, corresponds to a larger number of hydrocarbons detected (panel A of Fig.\,\ref{fig:C2H2trends}). This is very likely because the higher $F_{\rm ^{13}CCH_2}$/$F_{\rm C_2H_2}$ ratios correspond to large columns of gas being probed, or equivalently, higher gas optical depths. A large column of gas could allow sufficient column densities of less abundant hydrocarbons to be present in the line of sight such that their emission is above the noise level. This can explain the strong correlations between the detection rates of less dominant hydrocarbons such as \ch{C_3H_4} and \ch{C_2H_6} (Fig.\,\ref{fig:corr_zoom} and \ref{fig:correlations}). However, J1605 appears to be an outlier, which has a large $F_{\rm ^{13}CCH_2}$/$F_{\rm C_2H_2}$ ratio but only about half the number of organic molecules as detected in other sources with similar flux ratios. Furthermore, when $F_{\rm ^{13}CCH_2}$/$F_{\rm C_2H_2}$$>$0.3 the only inorganic molecule detected is \ch{CO_2} (and \ch{^{13}CO_2}), while in J1605 \ch{CO}, \ch{H_2O}, \ch{H_2}, and \ch{H \textsc{i}} have been detected in addition to \ch{CO_2}.

\subsubsection{Dust strengths}
Generally, the dust opacity limits the depth that can be probed in disks at infrared wavelengths. Due to efficient grain growth, VLMS disks are expected to have larger dust grain sizes \citep{2005Sci...310..834A} and consequently lower infrared dust opacities than their higher mass counterparts. The lower dust opacities can allow infrared emission to probe deeper layers of the disk and thus larger columns of gas (e.g. \citealt{2015A&A...582A.105A}). The left and middle column of panels in Fig.\,\ref{fig:features_summary} show the spectral appearance of the dust and the \ch{C_2H_2} features ordered by the $F_{\rm ^{13}CCH_2}$/$F_{\rm C_2H_2}$ ratio. We observe a general trend of higher gas optical depths probed in sources with lower dust strengths (F9.8, the continuum-normalized dust strength at 9.8\,$\mu$m; panel B of Fig.\,\ref{fig:C2H2trends}). However, there is some scatter. For example, NC1 has the strongest dust feature, but does not necessarily have the smallest flux ratio. This could be due to several factors, e.g., the infrared dust and molecular emission may arise from slightly different disk regions.

\subsubsection{Water and \ch{CO_2}}
As discussed in Sect.\,\ref{sec:detections}, the water detections anti-correlate with some of the organic molecules (Fig.\,\ref{fig:corr_zoom}). $F_{\rm CO_2}$ is the integrated line flux of \ch{CO_2} between 14.920\,$\mu$m - 14.993\,$\mu$m and $F_{\rm H_2O}$ is the sum of the integrated line fluxes of three water lines at 16.66\,$\mu$m, 17.10\,$\mu$m, and 17.36\,$\mu$m (16.654 - 16.673\,$\mu$m, 17.092 - 17.112\,$\mu$m, 17.354 - 17.363\,$\mu$m respectively). While there is likely underlying hydrocarbon emission, we do not identify strong hydrocarbon $Q-$branch emission at these wavelengths. Thus, non-detections provide upper limits on \ch{H_2O} line fluxes. At even longer wavelengths, the VLMS disks are very faint, and the sensitivity of MIRI also decreases, leading to high noise levels. 

In general, we observe that for sources in which we probe deeper columns of gas, the \ch{CO_2} and \ch{H_2O} emissions are weaker relative to \ch{C_2H_2} (panels C and D of Fig.\,\ref{fig:C2H2trends}, respectively). Since \ch{CO_2} is detected in more sources than \ch{H_2O}, this trend is more evident in panel C. The large scatter in panel D is likely due to the underlying hydrocarbon emission at the wavelengths where the water flux is measured. The $Q$-branch emission of \ch{CO_2} in panel C is less affected by the hydrocarbon emission as demonstrated by \citet{2025ApJ...984L..62A}. The flux ratios for the source J1558 are outliers. This is because J1558 is distinctively bright in \ch{C_2H_2} (as bright as in J1605, see Fig.\,\ref{fig:features_summary}) but hydrocarbon-poor unlike most of the sources in our sample (see Sect.\,\ref{sec:spectralappearance}). The observed trend of relative decrease in \ch{CO_2} and \ch{H_2O} fluxes could be due to i) lower abundances of these oxygen-bearing molecules, or ii) the hydrocarbon fluxes are more sensitive to increasing column density than \ch{CO_2} and \ch{H_2O} (see Sect.\,\ref{sec:discussion} and \citealt{2025ApJ...984L..62A}).

\subsection{Trends in stellar and disk properties}
\label{sec:stellardisktrends}

\begin{figure}
    \centering
    \includegraphics[width=0.65\linewidth]{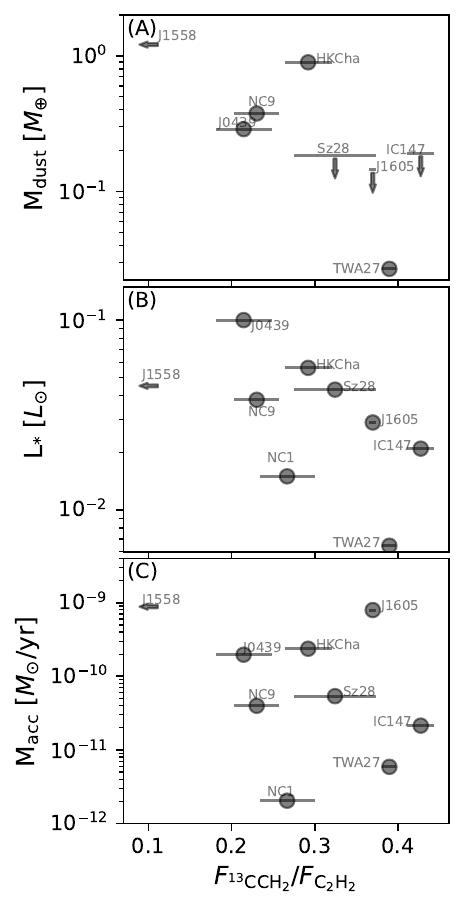}
    \caption{Trends in stellar and disk properties. Panels A\hl{-C} show the dust mass \hl{(estimated from 0.89\,mm continuum fluxes)}, stellar luminosity, and mass accretion rate against the flux ratio $F_{\rm ^{13}CCH_2}$/$F_{\rm C_2H_2}$. }
    \label{fig:stellardisktrends}
\end{figure}
While the mid-infrared wavelengths probe the inner disks, millimeter wavelengths probe the outer disk. Since dynamic processes, such as radial material transport, affect the inner disk as well as the outer disk, we compare the \ch{^{13}CCH_2}-\ch{C_2H_2} flux ratios of the inner disks with ALMA continuum fluxes which probe the outer disk dust masses. We also investigate relations with the stellar luminosity and the mass accretion rates.

Panel A of Fig.\,\ref{fig:stellardisktrends} show \hl{the dust masses calculated from ALMA 0.89\,mm fluxes (from \citealt{2023ASPC..534..539M}, \citealt{2017AJ....154...24R}) using the optically thin dust prescription of \citet{2023ASPC..534..539M} (Table\,\ref{tab:sources})}. The dust mass decrease with higher flux ratios indicating that disks with larger columns of hydrocarbons have lower pebble masses in the outer disk. A similar trend is also observed with stellar luminosity (panel B) where the gas optical depth probed by $F_{\rm ^{13}CCH_2}$/$F_{\rm C_2H_2}$ ratio increases with lower stellar luminosity. \citet{2025ApJ...984L..62A} show that the observed peak-to-continuum ratio of \ch{C_2H_2} increases with decreasing stellar luminosity. These findings point to a larger column of gas probed in disks around less luminous objects. This is in-line with non-detections of \ch{^{13}CCH_2} in all T\,Tauri disk spectra published so far (e.g. \citealt{Grant2023}), except one \citep{2024ApJ...977..173C}. No clear trend is observed with mass accretion rates (panel C). 

\begin{figure}[!hb]
    \centering
    \includegraphics[width=\linewidth]{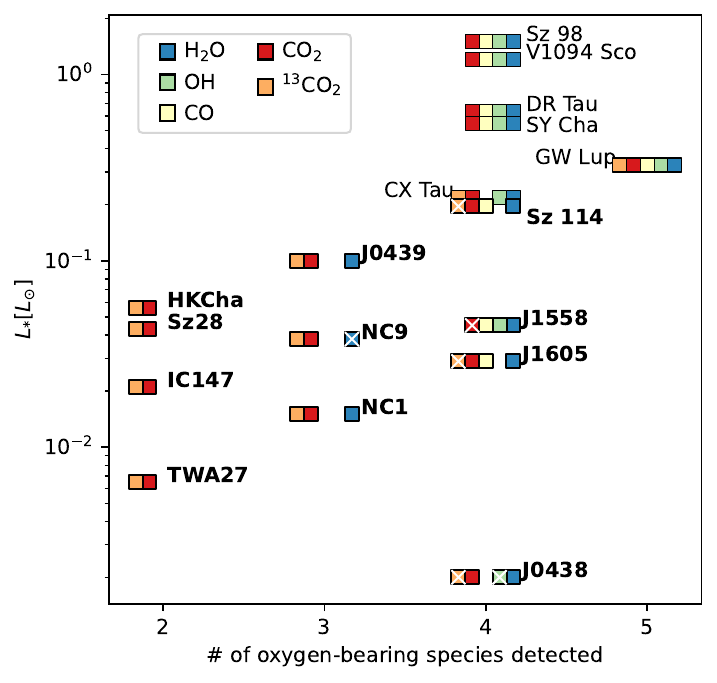}
    \caption{Luminosities of the central object against oxygen bearing molecules in the inner disks. The colored boxes refer to the oxygen-bearing molecules detected in the MIRI spectra of each source, indicated in the legend. Tentative detections are highlighted by white crosses. VLMS disks are highlighted in bold. The detection in T\,Tauri disks are taken from the following papers: GW\,Lup \citep{Grant2023}, Sz\,98 \citep{2023A&A...679A.117G}, DR\,Tau \citep{Temmink2024}, SY\,Cha \citep{2024ApJ...962....8S}, CX\,Tau \citep{2025A&A...693A.278V}, and V1094Sco (Tabone et al. in prep.). Detections of Sz114 are obtained from \citet{2023ApJ...959L..25X}.}
    \label{fig:Macc}
\end{figure}

\subsection{Oxygen carriers at mid-infrared wavelengths}
\label{sec:oxycarriers}
Figure\,\ref{fig:Macc} shows the stellar luminosity against the number of oxygen-bearing molecules observed in the inner disks. For comparison, we include six T\,Tauri disks from the MINDS sample (GW\,Lup, Sz\,98, DR\,Tau, SY\,Cha, CX\,Tau, and V1094Sco) and one other VLMS disk (Sz 114). In general, the more luminous objects, and thus the T\,Tauri disks, show a larger diversity of oxygen-bearing molecules compared to less luminous objects. While \ch{H_2O}, \ch{OH}, \ch{CO}, and \ch{CO_2} are the most commonly detected oxygen-bearing molecules in the T\,Tauri disks, \ch{^{13}CO_2} is detected only in sources known to be \ch{CO_2}-bright (i.e., in sources with weaker \ch{H_2O} emission), for example see \citet{Grant2023} and \citet{2025A&A...693A.278V}. On the other hand, in the VLMS disk sample, \ch{CO_2} and \ch{^{13}CO_2} are the most commonly detected oxygen-bearing molecules, while \ch{CO} emission is observed only in the two UpperSco objects. Although the VLMS disks generally have lower detection rates of oxygen-bearing molecules compared to the T\,Tauri disks, all sources show emission from at least two oxygen-bearing molecules (including the isotopologues). See \citet{2025ApJ...984L..62A} for a more detailed discussion on oxygen-bearing molecules, particularly water, in VLMS disks. See Grant et al. (in prep.) for a more detailed comparison of VLMS and T Tauri disk samples.

\section{Discussion}
\label{sec:discussion}
The evolution path towards an enhanced C/O gas composition can follow rapid inward material transport, carbon grain destruction, or \ch{H_2O}-ice trapping in the outer disk. These scenarios leave distinct signatures on the gas and dust compositions that can be probed at infrared and millimeter wavelengths. 

In the rapid inward material transport scenario, \citet{2023A&A...677L...7M} predict that the volatile C/O ratio would decrease initially (within $\sim$0.5\,Myr). Later, the C/O ratio would rapidly increase with the age of the disk, and that this enhancement is faster in disks around less massive stars than their higher-mass counterparts. Dynamic disk models such as \citet{2013A&A...554A..95P} suggest that VLMS disks experience rapid grain growth along with rapid inward material transport. 

Carbon grain destruction is also expected to enhance the C/O ratio in the disk. However, it can be expected to occur even in disks around higher-mass stars and is not clear why this process would be limited to only VLMS disks. This process does not necessarily require rapid inward material transport but can occur simultaneously and further enhance the C/O ratio.

\citet{2023ApJ...954...66K} and \citet{2024A&A...686L..17M} show that, in T\,Tauri disks, deep gaps can limit the delivery of oxygen-bearing ices to the inner disk and lead to a high C/O ratio in the inner disk within a few Myrs. Trapping of material in the outer disk could result in brighter outer disks compared to disks with efficient radial dust transport. In line with this, Kanwar et al. (subm.) show that a thermochemical model with a gap close to the \ch{H_2O} ice line that separates an oxygen-depleted (high C/O) inner disk and normal outer disk can reproduce the MIRI spectrum of the hydrocarbon-rich disk of J1605.

\subsection{Tentative evidence for C/O enhancement with age}
The evolution timescales of the VLMS disks are expected to be short. Moreover, during this evolution, the stellar luminosities are also expected to decrease. The age of the individual objects are not available, but we find that sources with lower stellar luminosities show higher $F_{\rm ^{13}CCH_2}$/$F_{\rm C_2H_2}$ ratio (and are possibly more carbon-rich, Fig.\,\ref{fig:stellardisktrends}).

\begin{figure}
    \centering
    \includegraphics[width=\linewidth]{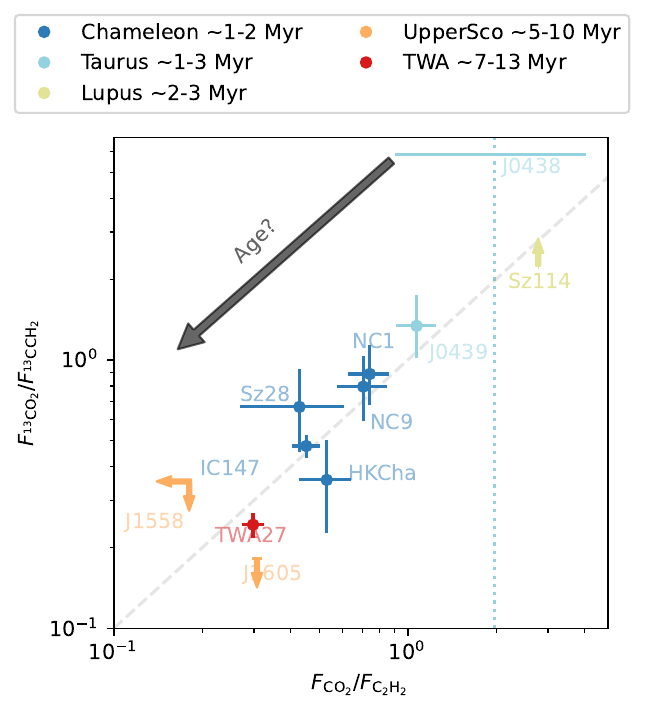}
    \caption{Variation of flux ratios $F_{\rm CO_2}/F_{\rm C_2H_2}$ and $F_{\rm ^{13}CO_2}/F_{\rm ^{13}CCH_2}$ across different star-forming regions or moving group. We include Sz114 \citep{2023ApJ...959L..25X}. The dashed line indicates a line with slope of unity. Due to the non-detections of both \ch{^{13}CO_2} and \ch{^{13}CCH_2} in J0438, we present it with a vertical dotted line. The colors indicate the different star-forming regions or the moving group shown in the legend at the top. \hll{The ages indicated are canonical values commonly used in the literature; however, recent studies indicate a larger diversity in the ages of sources within each star-forming region (e.g., \citealt{2023A&A...678A..71R}).}}
    \label{fig:SFR}
\end{figure}

Although the age of individual objects is not well constrained, the typical age of their star-forming regions or moving groups can provide some insights into the evolution of these objects. Figure\,\ref{fig:SFR} shows the flux ratio of the isotopologues of the predominant oxygen and carbon carriers in the spectra ($F_{\rm ^{13}CO_2}$/$F_{\rm ^{13}CCH_2}$) versus the flux ratio of the predominant carriers themselves ($F_{\rm CO_2}$/$F_{\rm C_2H_2}$). The trend is very close to the line with a slope of unity in the log-log scale. This can be explained by the ISM fractionation ratio of \ch{^{12}C}/\ch{^{13}C}$\sim$70 \citep{2009ApJ...693.1360W}; neglecting the optical depth effects, $F_{\rm ^{13}CO_2}/F_{\rm ^{13}CCH_2}=0.5\cdot F_{\rm CO_2}/F_{\rm C_2H_2}$, or $log_{10}\left(F_{\rm ^{13}CO_2}/F_{\rm ^{13}CCH_2}\right)=log_{10}\left(0.5\right)+ log_{10}\left(F_{\rm CO_2}/F_{\rm C_2H_2}\right)$. We do not capture the offset of $log_{10}\left(0.5\right)$  in Fig.\,\ref{fig:SFR} because we show here only the integrated fluxes around the peaks and not the entire molecular flux. Interestingly, the sources of the same star-forming regions (indicated by the colors shown in the legend, Table\,\ref{tab:sources}) lie closely grouped. We have included Sz114 \citep{2023ApJ...959L..25X} for completeness. The objects in an older star-forming region or a moving group, i.e., UpperSco and TWA, are at the bottom left. Following the dashed line, we find Chameleon I sources and then the sources from Taurus and Lupus with higher flux ratios. Although we indicate the traditionally assumed ages of the star-forming regions, a recent study of the Sco-Cen association has shown that the ages of the star-forming regions are much more intricate and diverse, with different parts of the same star-forming regions possibly having different ages \citep{2023A&A...678A..71R}. While we indicate the age of the Chameleon I region as 1-2\,Myr, the study suggests an age of $\sim$4\,Myr, and for the UpperSco and TWA objects ages much larger than 7\,Myr. These ages support the trend of decrease in the \ch{CO_2} to \ch{C_2H_2} flux ratios with age.

Models such as \citet{2023A&A...677L...7M} show that while the T\,Tauri disks remain oxygen-rich beyond 8\,Myr, due the short viscous timescales in VLMS disks, their inner
disks become carbon-rich in a few Myrs (e.g. 2\,Myr for a star of mass 0.1\,$M_{\odot}$ even using a low viscosity of $\alpha$=$10^{-4}$). These models therefore predict that older disks should in general show enhanced carbon-rich gas composition. This is in-line with the trend observed in Fig.\,\ref{fig:SFR}. 

\hl{However, there are some important caveats to consider. The ages of both the star-forming regions and the individual objects within them are subject to significant uncertainties. Furthermore, environmental differences between regions may substantially influence the observed trends. For example, UpperSco is exposed to elevated UV radiation fields due to the presence of nearby early-type massive stars, which can affect disk evolution. Additionally, we need a larger sample of sources in each star-forming region to confirm the observed trend. }

J1558 and J1605 are nearly identical in terms of their stellar parameters such as luminosity, mass, and accretion rates (see Table\,\ref{tab:sources}), and belong to the UpperSco star-forming region. Yet, the spectral appearances and gas composition are strikingly distinct (Fig.\,\ref{fig:detectionrates}, \ref{fig:C2H2trends}). There could still be sources in the same star-forming regions with different evolutionary pathways due to different disk substructures \citep{2023ApJ...959L..25X}, for example due to planets carving gaps. Moreover, switching the ratios on either axes no trend was observed between $F_{\rm ^{13}CO_2}/F_{\rm CO_2}$ and $F_{\rm ^{13}CCH_2}/F_{\rm C_2H_2}$. This indicates that the \ch{CO_2} column density likely does not decrease with larger \ch{C_2H_2} column densities or higher C/O ratios, e.g. \citet{2024A&A...689A.231K} showed that sufficient \ch{CO_2} can form even in high C/O conditions. It could also be that disk substructures play a major role. Kanwar et al. (subm.) show that in a VLMS disk model with a gap, \ch{C_2H_2} emits from the inner disk depleted of oxygen and dust, while \ch{CO_2} emits from the outer disk which has a normal C/O and dust-to-gas mass ratio. \hl{More observations and disk modelling are required to validate these tentative trends.}

\subsection{Signatures of disk evolution in inner and outer disk dust}
\begin{figure*}[!ht]
    \centering
    \includegraphics[width=0.75\linewidth]{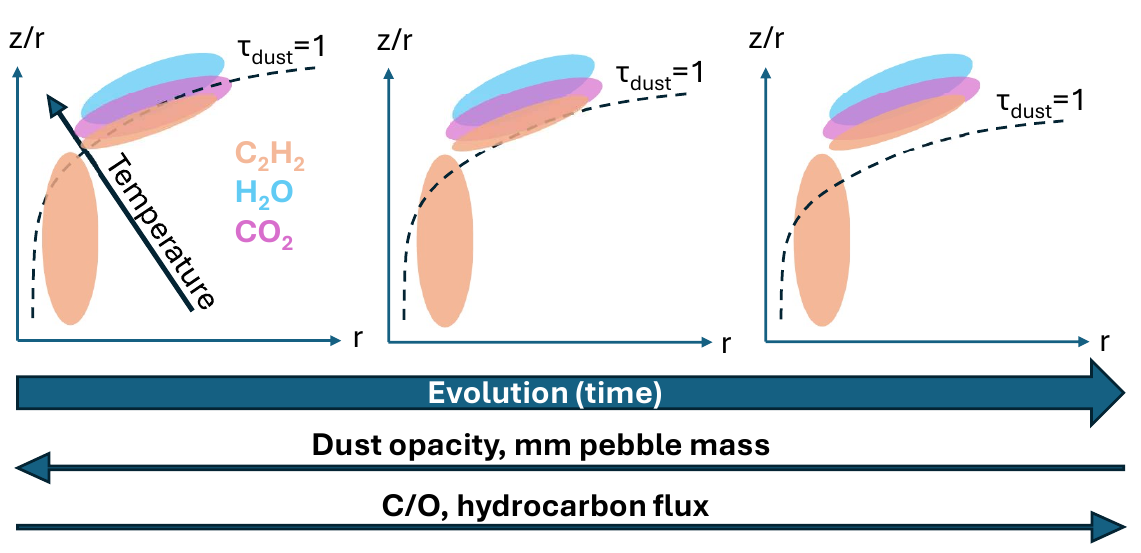}
    \caption{Cartoon illustrating the role of dust opacity on the mid-IR spectral appearance. The dotted line represents the $\tau_{\rm dust}$=1 layer. Cyan, pink, and orange ellipses represent the \ch{H_2O}, \ch{CO_2}, and the two \ch{C_2H_2} reservoirs. The arrows show the direction of increase in the quantity mentioned.}
    \label{fig:evo}
\end{figure*}

While transport processes are suggested to enhance the C/O in the inner disk \citep{2023A&A...677L...7M}, efficient grain growth and dust settling as predicted by \citet{2013A&A...554A..95P} would lead to a decrease in the number density of small dust in the surface layers that largely carry the continuum opacity in the infrared wavelengths. Thus, grain growth and settling would lead to weaker dust features and allow probing deeper layers of the disk in the infrared (in other words, large columns of gas). In fact, this is what we find in Sect.\,\ref{sec:miritrends}, i.e., disks with larger gas columns ($F_{\rm ^{13}CCH_2}/F_{\rm C_2H_2}$) show weaker dust strengths (F9.8).

Further, rapid inward transport of material would deplete the outer disk pebbles leading to smaller disk masses \citep{2020A&A...638A..88L} and lower continuum fluxes at millimeter wavelengths for evolved disks. In Sect.\,\ref{sec:stellardisktrends} we do find a trend that disks with higher \ch{C_2H_2} gas columns have lower disk dust masses.

The observed trends in millimeter and infrared dust signatures are in-line with models that predict high C/O by transport processes \citep{2023A&A...677L...7M}. But, this does not completely rule out the possibility of C/O enhancement by carbon grain destruction or dust traps locking up \ch{H_2O} ice. Moreover, the above comparisons are qualitative, and a more quantitative comparison between such models and observed data is required.

\subsection{Spectral appearance and the disk structure}
In Sect.\,\ref{sec:miritrends}, we show that sources with higher gas columns ($F_{\rm ^{13}CCH_2}/F_{\rm C_2H_2}$) show weaker \ch{CO_2} and \ch{H_2O} fluxes relative to \ch{C_2H_2}. Although VLMS disks generally show less diverse oxygen-bearing molecules than T\,Tauri disks (Sect.\,\ref{sec:oxycarriers}), the VLMS disks more commonly show the presence of \ch{^{13}CO_2}, possibly indicating a larger observable column of \ch{CO_2} than typical T\,Tauri disks. As shown by \citet{2025ApJ...984L..62A}, lower detection rates of oxygen-bearing molecules in the carbon-rich VLMS disks can simply be due to the emission from \ch{C_2H_2} and other hydrocarbons outshining the emission from the oxygen-bearing molecules. In general, the spectral appearance of the disk would be strongly influenced by two important factors, the spatial distribution of different chemical species and the continuum optical depth $\tau_{\rm dust}$=1 surface. These are in turn related to the grain growth and C/O enhancement timescales, and the inner disk substructure. 

\subsubsection{Role of dust opacity}
Figure\,\ref{fig:evo} shows a possible evolution pathway for the $\tau_{\rm dust}$=1 surface that could explain the observed trends in Sect.\,\ref{sec:miritrends}. \citet{2018A&A...618A..57W} showed that the emission from different molecules arises from onion-like layers at different depths (at solar C/O ratio). In such a scenario, changes in dust opacities can expose different chemical layers. We neglect the feedback that the change in opacities can have on chemistry, which can be important. As the disk evolves, the lower dust opacity due to grain growth shifts the $\tau_{\rm dust}$=1 layer deeper in the disk. Following the structure in \citet{2018A&A...618A..57W}, \ch{H_2O} is at the top, followed by \ch{CO_2} and even deeper is the \ch{C_2H_2} emitting layer. In addition, \citet{2024A&A...689A.231K} show that \ch{C_2H_2}, and in general the hydrocarbons, have two reservoirs, one in the surface layer and the other much closer to the midplane.

When the $\tau_{\rm dust}$=1 layer is high up in the disk, only the gas in the warm surface layers emits. The spectra would appear oxygen-rich (dominated by water) similar to T\,Tauri disks as a result of weak emission from \ch{C_2H_2} and other hydrocarbons due to their low column densities above the $\tau_{\rm dust}$=1 layer. Any \ch{C_2H_2} emission would be weak but warm; a good example of this would be the MIRI spectrum of Sz114 \citep{2023ApJ...959L..25X}. 

A slightly lower $\tau$=1 layer would lead to much brighter but optically thin (i.e., no molecular continuum) \ch{C_2H_2} emission. The brightness of \ch{C_2H_2}, \ch{CO_2}, and \ch{H_2O} would be related to the radial extent of these species. Gaps and cavities can have a strong influence on the flux levels. Depending on the radial extent and abundance of \ch{C_2H_2} and \ch{H_2O} in the surface layers, the spectral appearance could vary from \ch{C_2H_2}-bright (e.g. J1558) to \ch{H_2O}-bright (e.g. J0438). 

An even deeper $\tau$=1 layer would allow gas emission from layers increasingly closer to the midplane. In such cases, while \ch{H_2O} and \ch{CO_2} would still emit from the surface layers, a deeper reservoir of hydrocarbons would now emit much stronger than \ch{H_2O}. \citet{2025ApJ...984L..62A} show that there is a strong change in the peak-to-continuum ratio of \ch{C_2H_2} in the sample, and a weak change of \ch{CO_2}, but not for water. \ch{CO_2} would still be visible due to the prominent $Q-$branch (see Fig.\,6 of \citealt{2025ApJ...984L..62A}) and \ch{^{13}CO_2} would be visible due to the large column density of \ch{CO_2} above the $\tau$=1 layer. Deeper gas columns would reflect lower \ch{C_2H_2} temperatures compared to when the $\tau$=1 surface is higher up in the disk. The trend observed in the diversity of oxygen-carriers (Fig.\,\ref{fig:Macc}) could simply be due to the strong hydrocarbon emission in disks of lower stellar luminosities outshining the oxygen-bearing species.

\hl{Although we neglected the effects of changing dust opacity on the temperature and chemistry of the gas, a lower dust opacity could allow more energy to be absorbed by the gas compared to cases with higher opacity. This may result in elevated gas temperatures \citep[e.g. see][]{2019A&A...631A..81G}. Deeper penetration of UV radiation can strongly influence the chemical composition of the disk \citep[e.g. see][]{2019A&A...631A..81G}, but large columns of molecules can shield themselves and other molecules \citep[e.g.][]{2022ApJ...930L..26B,2024A&A...683A.219W}. To accurately assess the influence of the change in dust opacity on gas temperature and chemistry, and the validity of the scenarios presented in Fig.\,\ref{fig:evo}, detailed modeling that includes molecular self-shielding—especially among the many hydrocarbons—is essential. Another important factor affecting the temperatures is the luminosity of the central objects, which generally decreases with age \citep[e.g.][]{1986ApJ...311..226N}; moreover, stellar activity can also have a strong impact on the thermal structure and chemical evolution of the disk.}

\subsubsection{Role of C/O ratio}
The strong spectral signatures of hydrocarbons and weaker dust opacity would still require a hydrocarbon reservoir with a C/O ratio greater than unity (\citealt{2024A&A...689A.231K}, \citealt{2018A&A...618A..57W}). However, \citet{2024A&A...689A.231K} show that even in an enhanced C/O environment (as high as C/O=2), large column densities ($>10^{18}$\,cm$^{-2}$) of \ch{H_2O} and \ch{CO_2} can still be present. \hll{This is in line with the analysis of water in our sample \citep{2025ApJ...984L..62A}.}

\hl{\citet{2018A&A...614A...1W} show that condensation of elements, starting from solar abundances, could enhance the gas phase C/O ratio, but only up to about 0.83 at temperatures of $\sim$300\,K. This suggests that more extreme processes are required to increase the gas phase C/O ratio to beyond unity.} 

Such enhancements are more likely in a VLMS disk than for a T\,Tauri disk \citep{2023A&A...677L...7M}. i.e., rapid transport processes could enhance the C/O in the inner disk, while grain growth would allow probing much deeper into the disk. These two processes together could render the spectral appearance of evolved disk appear very carbon-rich. Although dust growth and inward transport can also happen in a T\,Tauri disk they are less efficient than in VLMS disks (\citealt{2013A&A...554A..95P}, \citealt{2023A&A...677L...7M}). 

Rapid inward transport of material in the VLMS disks would lead to an oxygen-depleted disk with a high C/O. However, C/O enhancement can also occur through gas phase carbon-enrichment by the destruction of carbon in the dust grains \citep{2023NatAs...7..805T} or trapping oxygen-rich ices in the outer disk \citep{2024A&A...686L..17M}. \citet{2024Sci...384.1086A} point out that oxygen-depletion and carbon-enrichment scenarios could lead to measurable differences in the gas and dust compositions of the disk. While the observed trends in the infrared and millimeter wavelengths favor the rapid transport scenario, the carbon enrichment by carbon grain destruction can occur simultaneously or even be the main C/O enriching process in the midplane closer to the star. 

\hll{The efficiency of transport processes vary with height above the disk midplane, potentially leading to vertical stratification in the C/O ratio. Additionally, steep vertical gradients in gas temperatures and UV radiation in the disk can contribute to stratified carbon grain destruction. Whether such stratification persists over the disk evolution timescales depends on the strength of vertical mixing. Efficient vertical mixing acts to remove this vertical stratification of elemental abundances \citep{2022A&A...668A.164W}. This mixing can also change the molecular abundances in the layers probed in the mid-infrared \citep{2011ApJ...731..115H,2022A&A...668A.164W}. For example, \citet{2022A&A...668A.164W} found that stronger vertical mixing leads to richer hydrocarbon chemistry in the upper layers, leading to an increase in \ch{C_2H_2} fluxes of more than an order of magnitude. Importantly, this enhanced hydrocarbon emission does not necessarily indicate a carbon-rich environment, as water abundances are also elevated. As shown by \citet{2025ApJ...984L..62A}, once the hydrocarbon column densities exceed a critical threshold, they can outshine water emission$-$even when column densities of water are higher overall.}

Further analyses by retrieving the column densities, temperatures and emitting area of the gas emission, as well as the chemical composition of the dust grains, or by full forward modeling (e.g., \citealt{2024A&A...683A.219W}) of the entire MIRI spectrum can provide strong constraints on the C/O enhancement processes. Kanwar et al. (subm.) show that using multiple ($>$2) molecular emissions, it might be possible to distinguish between the C/O enhancement processes. 

Robust spatially resolved ALMA observations would provide valuable insight into the importance of the disk substructures that can trap oxygen-rich material in the outer disk. It is important to note here that all of our sources were selected based on the rationale that they were previously studied with \textit{Spitzer}. This could introduce a bias towards brighter VLMS targets. A larger sample of VLMS sources would further test our findings and expand our understanding of disk evolution. 

Finally, since some of the C/O enrichment processes could occur in the T Tauri disks, possibly at a different rate, a comparison study between VLMS and T Tauri disk spectra can provide valuable insights into the inner disk environments (e.g. Grant et al. in prep.). Parameter exploration using thermochemical models to understand the role of inner disk substructures and elemental composition in the mid-infrared spectra of the VLMS disks would be insightful in conjunction with these observations. Expanding dynamic models such as \citet{2023A&A...677L...7M} to 2D and deriving molecular mid-infrared spectra would provide a better understanding of the interplay between the dust evolution and chemical composition. 

\section{Conclusions}
\label{sec:conclusions}
We present the MIRI spectra of disks around ten very low-mass stars with the central objects ranging from 0.02\,$M_{\odot}$ to 0.14\,$M_{\odot}$, across three star-forming regions and one moving group. We draw the following conclusions based on the analysis of these spectra:
\begin{enumerate}
    \item The VLMS spectra show a diversity of dust features. In agreement with the \textit{Spitzer} results, these dust features are typically broader and weaker than in the disks around the higher-mass stars.
    \item The spectra are rich in molecular emission. We detect \ch{H\,\textsc{i}}, [\ch{Ar\,\textsc{ii}}], [\ch{Ne\,\textsc{ii}}], \ch{H_2}, \ch{OH}, \ch{H_2O}, \ch{CO}, \ch{CO_2}, \ch{HCN}, \ch{HC_3N}, \ch{CH_3}, \ch{CH_4}, \ch{C_2H_2}, \ch{C_2H_4}, \ch{C_2H_6}, \ch{C_3H_4}, \ch{C_4H_2}, \ch{C_6H_6}, \ch{^{13}CO_2} and \ch{^{13}CCH_2} in the sample. 
    \item Organic molecules dominate the molecular emission. The most commonly detected molecules are \ch{C_2H_2}, and \ch{HCN}, followed by \ch{CO_2}, \ch{C_4H_2}, and \ch{C_6H_6}.
    \item The detection rates of organic molecules correlate with other organic molecules while anti-correlating with the inorganic molecules.
    \item Spectra with a higher $F_{\rm ^{13}CCH_2}$/$F_{\rm C_2H_2}$ ratio (a proxy for column density of gas probed by MIRI) show detections of a large number of hydrocarbons. The ratio also anti-correlates with the 10\,$\mu$m dust strengths, disk dust mass, stellar luminosity, and the flux ratios of oxygen-bearing molecules to acetylene ($F_{\rm H_2O}$/$F_{\rm C_2H_2}$ and $F_{\rm CO_2}$/$F_{\rm C_2H_2}$).
    \item We find tentative evidence for the chemical evolution from oxygen-rich young disks to carbon-rich older disks. The anti-correlations with the 10\,$\mu$m dust strength and the disk dust mass suggest rapid inward material transport and grain growth in-line with model predictions such as \citet{2023A&A...677L...7M}.
\end{enumerate}
In this paper, we have analyzed the gas emission in the MINDS VLMS sample. We refer to Jang et al. (in prep.) for a detailed analysis of the dust in these disks and \citet{2025ApJ...984L..62A} for an overview of oxygen-bearing molecules in these disks.

\begin{acknowledgements}
This work is based on observations made with the NASA/ESA/CSA James Webb Space Telescope. The data were obtained from the Mikulski Archive for Space Telescopes at the Space Telescope Science Institute, which is operated by the Association of Universities for Research in Astronomy, Inc., under NASA contract NAS 5-03127 for JWST. These observations are associated with program \#1282. The following National and International Funding Agencies funded and supported the MIRI development: NASA; ESA; Belgian Science Policy Office (BELSPO); Centre Nationale d’Etudes Spatiales (CNES); Danish National Space Centre; Deutsches Zentrum fur Luft- und Raumfahrt (DLR); Enterprise Ireland; Ministerio De Econom\'ia y Competividad; Netherlands Research School for Astronomy (NOVA); Netherlands Organisation for Scientific Research (NWO); Science and Technology Facilities Council; Swiss Space Office; Swedish National Space Agency; and UK Space Agency. I.K., A.M.A., and E.v.D. acknowledge support from grant TOP-1 614.001.751 from the Dutch Research Council (NWO). A.C.G. acknowledges support from PRIN-MUR 2022 20228JPA3A “The path to star and planet formation in the JWST era (PATH)” funded by NextGeneration EU and by INAF-GoG 2022 “NIR-dark Accretion Outbursts in Massive Young stellar objects (NAOMY)” and Large Grant INAF 2022 “YSOs Outflows, Disks and Accretion: towards a global framework for the evolution of planet forming systems (YODA)”. G.P. gratefully acknowledges support from the Max Planck Society and from the Carlsberg Foundation, grant CF23-0481. E.v.D. acknowledges support from the ERC grant 101019751 MOLDISK and the Danish National Research Foundation through the Center of Excellence ``InterCat'' (DNRF150). T.H. and K.S. acknowledge support from the European Research Council under the Horizon 2020 Framework Program via the ERC Advanced Grant Origins 83 24 28. I.K. and J.K. acknowledge funding from H2020-MSCA-ITN-2019, grant no. 860470 (CHAMELEON). B.T. is a Laureate of the Paris Region fellowship program, which is supported by the Ile-de-France Region and has received funding under the Horizon 2020 innovation framework program and Marie Sklodowska-Curie grant agreement No. 945298. V.C. acknowledge funding from the Belgian F.R.S.-FNRS. D.G. thanks the Belgian Federal Science Policy Office (BELSPO) for the provision of financial support in the framework of the PRODEX Programme of the European Space Agency (ESA). D.B. and M.M.C. have been funded by Spanish MCIN/AEI/10.13039/501100011033 grants PID2019-107061GB-C61 and No. MDM-2017-0737. M.T., M.V. and A.D.S acknowledge support from the ERC grant 101019751 MOLDISK. I.P. acknowledges partial support by NASA under agreement No. 80NSSC21K0593 for the program ``Alien Earths.'' P.P. thanks the Swiss National Science
Foundation (SNSF) for financial support under grant number 200020\_200399. 
\end{acknowledgements}

%
  \bibliographystyle{aa} 
  \bibliography{ref.bib} 
%
\appendix
\onecolumn

\section{De-reddening and continuum definition}
\label{sec:continuum}
\begin{figure*}[!ht]
    \centering
    \includegraphics[width=0.85\linewidth]{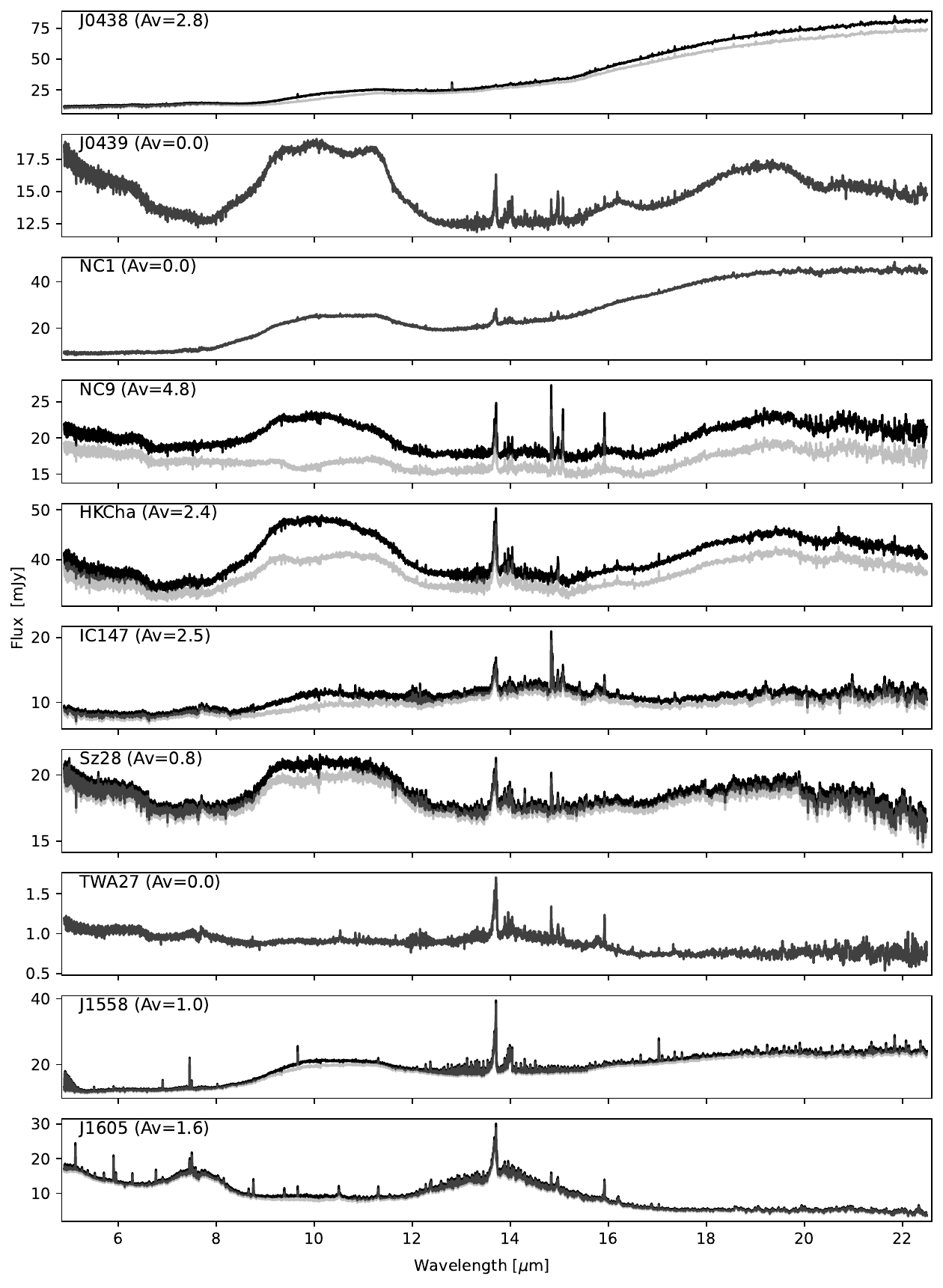}
    \caption{Spectra before (gray) and after (black) dereddening.}
    \label{fig:deredden}
\end{figure*}

Figure\,\ref{fig:deredden} shows the MIRI spectra before and after de-reddening. The extinction values and the references are listed in Table\,\ref{tab:sources}. The literature values of the extinctions are based on the extinction curve given by \citet{1989ApJ...345..245C} assuming $R_v$=3.1. However, this does not extend to the entire MIRI wavelength range. Hence, we use the extinction curve described in \citet{2023ApJ...950...86G}. The spectra are normalized to 150\,pc.

\begin{figure*}[!ht]
    \centering
    \includegraphics[width=0.85\linewidth]{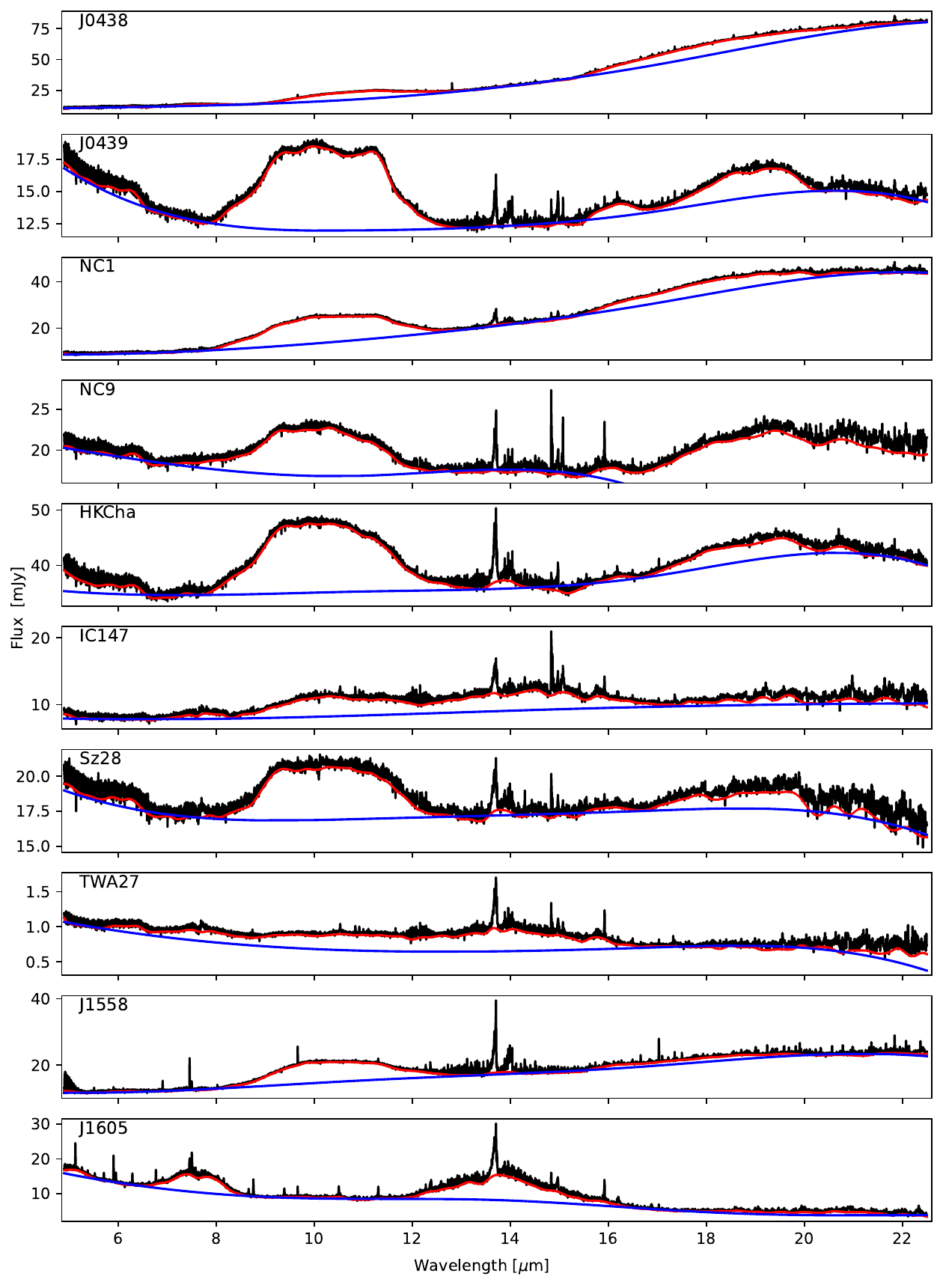}
    \caption{Continuum definition. The observed MIRI spectra are shown in black, the continua with narrow Savitzky-Golay filter are shown in red, and the continua with broad Savitzky-Golay filter are shown in blue.}
    \label{fig:continuum}
\end{figure*}

Figure\,\ref{fig:continuum} shows the continuum curves used in this paper. As explained in Sect.\ref{sec:datareduction}, we use a narrow and a broad Savitzky-Golay filter to determine two sets of continua for our sample (shown in red and blue in the figure respectively). The broad width continuum (blue) is only used to measure the dust strength and the line fluxes of \ch{C_2H_2}, and hence the shape of this continuum beyond 14\,$\mu$m or shortward of 8\,$\mu$m is not important. The narrow width continuum (red) is used to measure the line fluxes of all the other molecules used in the analysis of the paper. The spectra are normalised to 150\,pc.

\section{Comparison with \textit{Spitzer}}
\label{sec:spitzer}
\begin{figure*}[!ht]
    \centering
    \includegraphics[width=0.83\linewidth]{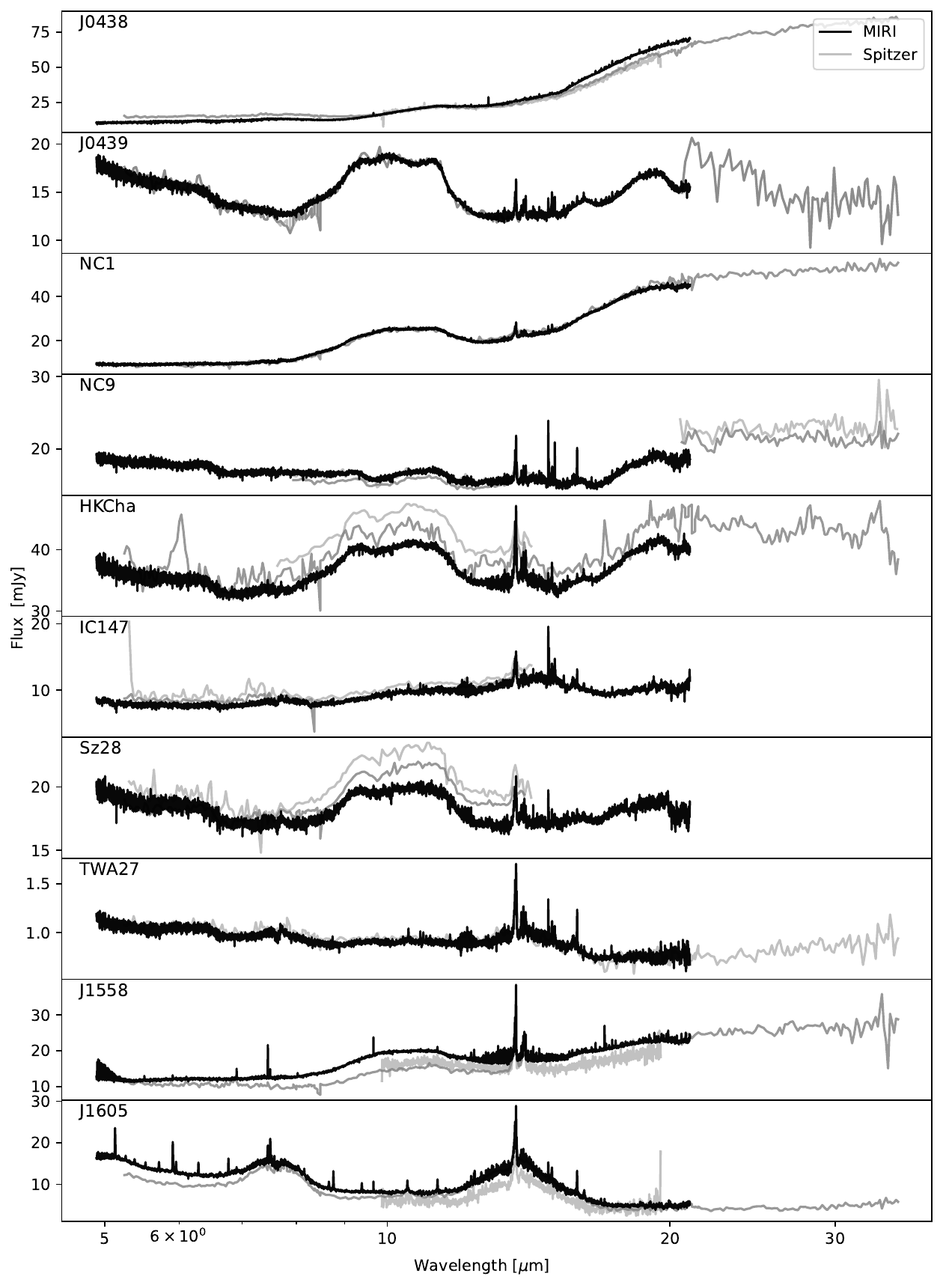}
    \caption{Comparison of the MIRI observations with \textit{Spitzer} spectra.}
    \label{fig:features_SED_zoomin}
\end{figure*}

Figure\,\ref{fig:features_SED_zoomin} compares the MIRI observations with the \textit{Spitzer} spectra (both normalized to 150\,pc). The \textit{Spitzer} specrta were taken from \citet{2009ApJ...696..143P}, \citet{2013ApJ...779..178P}, and CASSIS (\citealt{2011ApJS..196....8L}, \citealt{2015ApJS..218...21L}). In general the MIRI spectra seem to match well with the \textit{Spitzer} spectra. In some sources there seem to be some flux differences. For example, J0438 shows a seesaw variability (more details in \citealt{2025arXiv250411424P}). In some sources, e.g. Sz28, HKCha, and J1558, there is a flux discrepancy between MIRI and \textit{Spitzer}. This is more evident at the 10\,$\mu$m dust feature.

\clearpage

\section{Detection criteria}
\label{app:detections}
\begin{figure}[!ht]
    \centering
    \includegraphics[trim={5.5em 1em 7.5em 4em}, clip, width=0.995\linewidth]{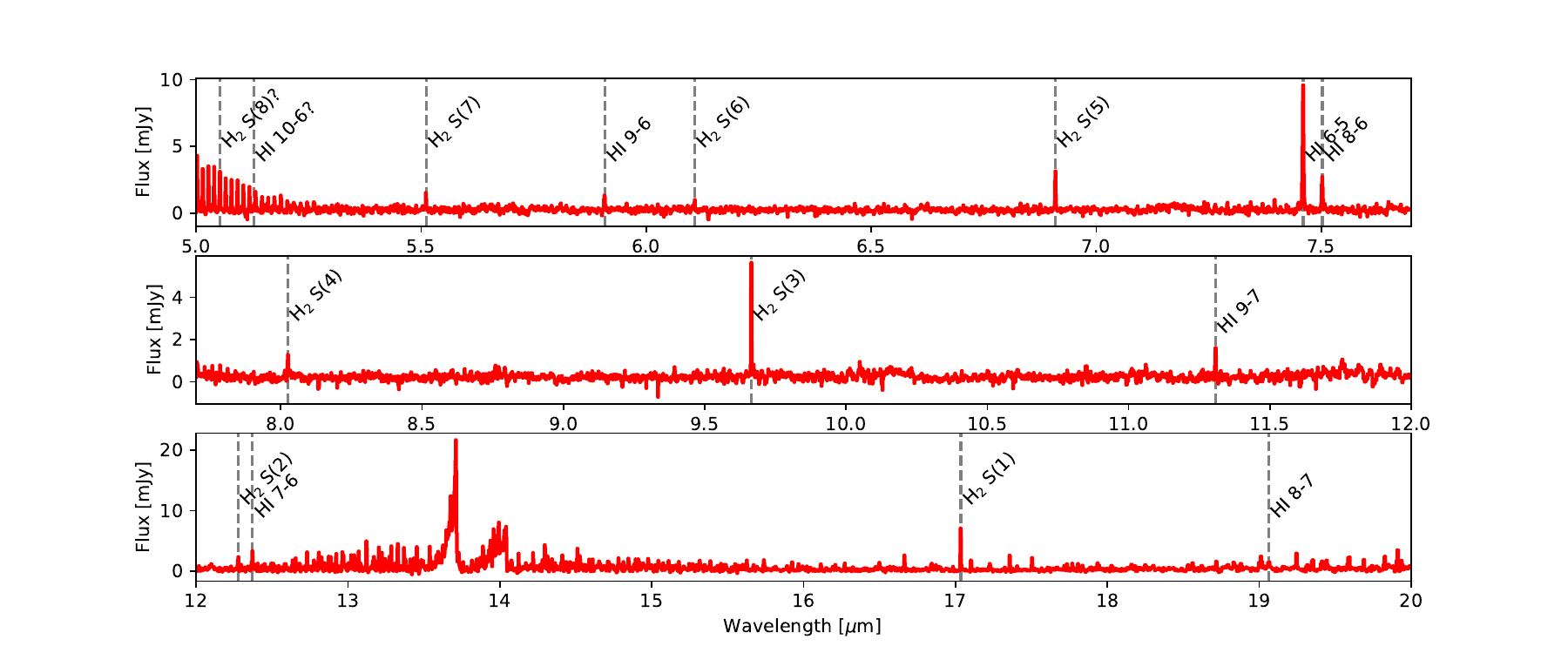}
    \caption{Detections of atomic and molecular hydrogen lines in the J1558 spectrum. The continuum-subtracted MIRI spectrum is shown in red. The atomic and molecular hydrogen lines are marked by vertical dashed lines and are labeled. The two lines with question marks indicate line emissions that could be possibly blended with CO ro-vibration lines.}
    \label{fig:j1558Hydrogen}
\end{figure}

\begin{figure}[!ht]
    \centering
    \includegraphics[trim={0 0 0 0.5em}, clip, width=0.48\linewidth]{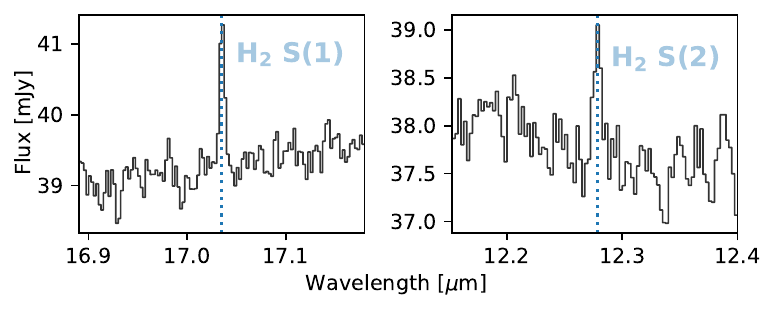}
    \caption{Detections of molecular hydrogen line emission in the HKCha spectrum. The MIRI spectrum is shown in black.}
    \label{fig:hkcha_h2}
\end{figure}

\begin{figure}[!ht]
    \centering
    \includegraphics[width=0.95\linewidth]{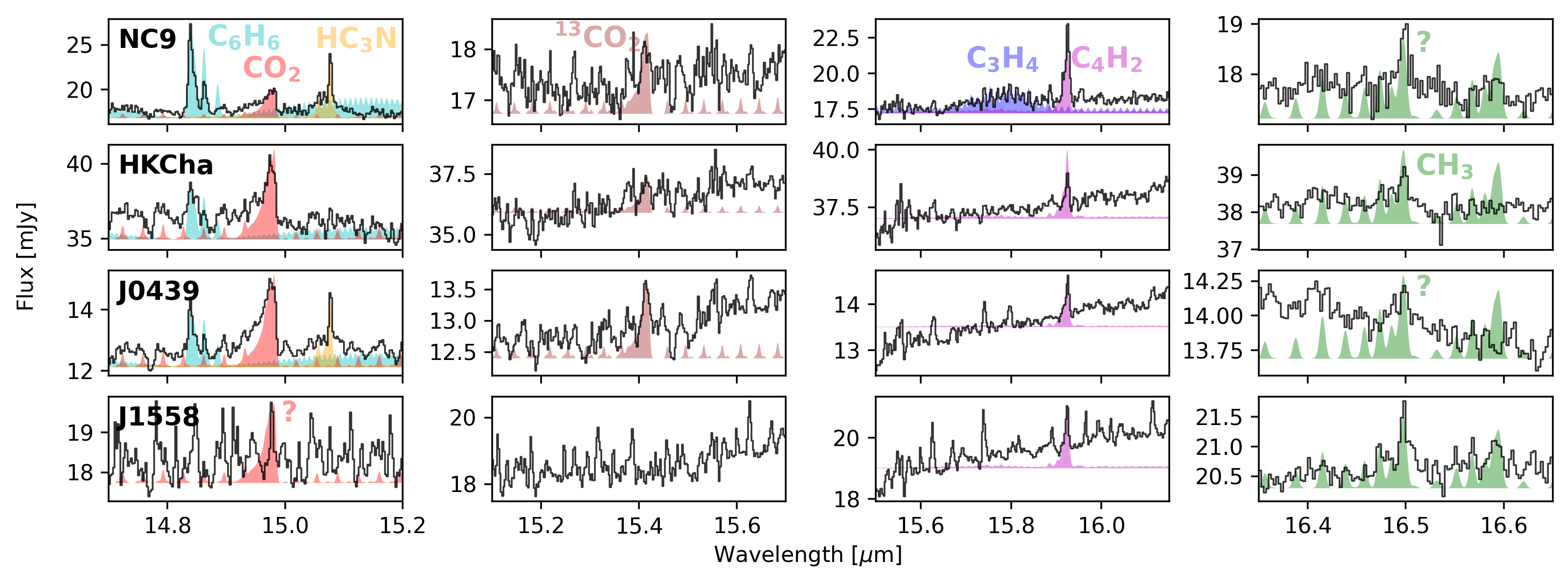}
    \caption{Detections of molecular emission in NC9, HKCha, J0439, and J1558. The MIRI spectra are shown in black. The different molecular emissions are shown using slab model spectra as colored filled regions in each panel. `?' indicates more modeling is required to confirm the detection.}
    \label{fig:remaindetect}
\end{figure}

\begin{figure}[!ht]
    \centering
    \includegraphics[width=0.7\linewidth]{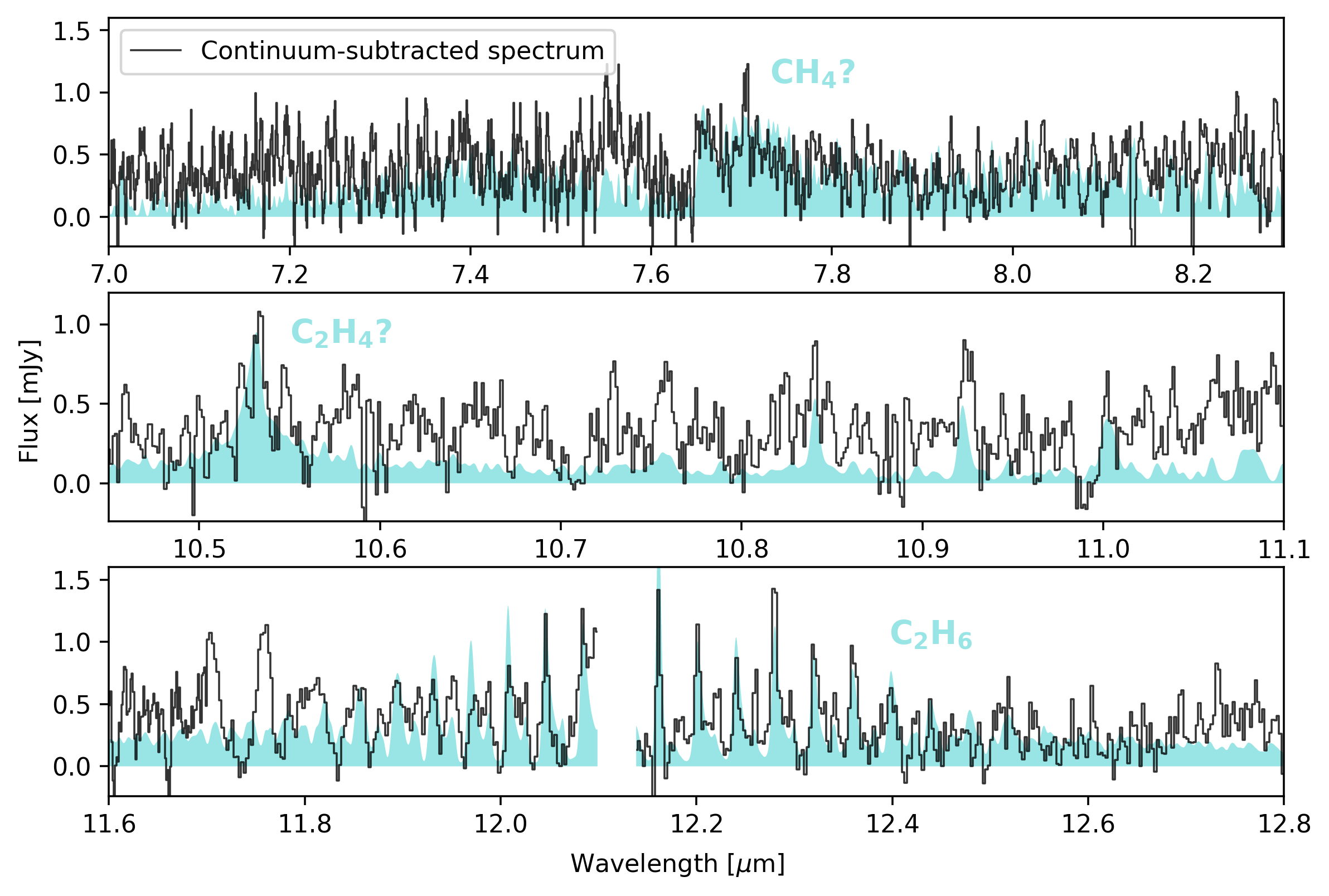}
    \caption{Detections of molecular emission in NC9. The continuum-subtracted MIRI spectrum is shown in black. The different molecular emissions are shown using slab model spectra as colored filled regions in each panel. `?' indicates more modeling is required to confirm the detection.}
    \label{fig:nc9_c2h6}
\end{figure}

\begin{figure}[!ht]
    \centering
    \includegraphics[width=0.45\linewidth]{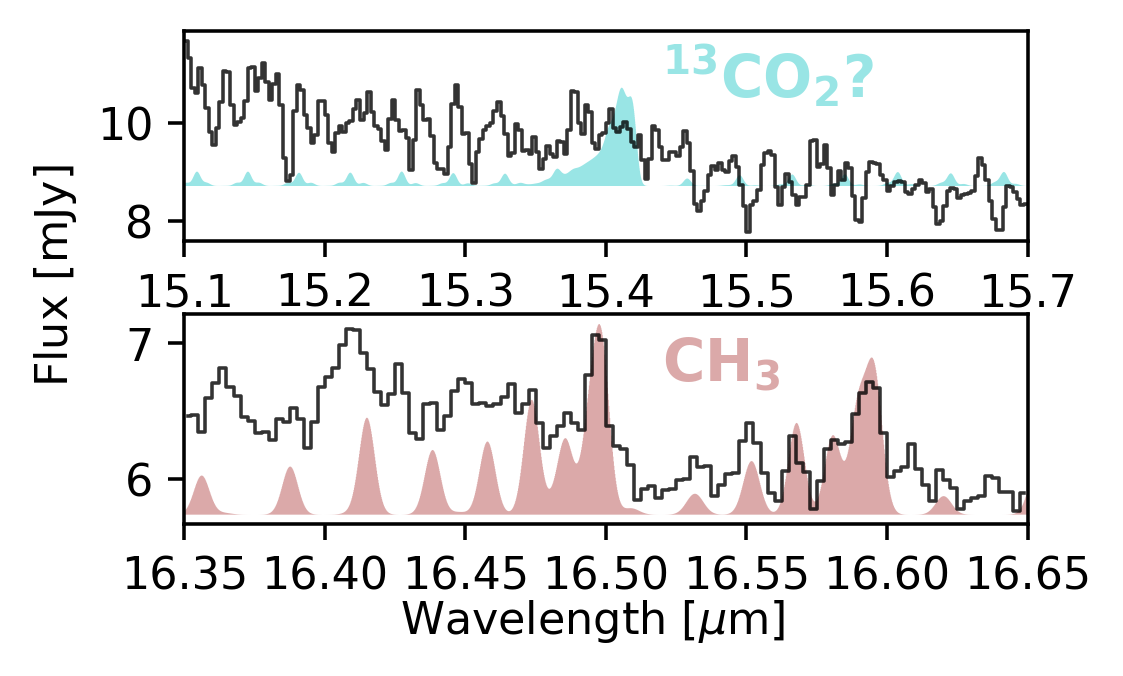}
    \caption{Detections of molecular emission in J1605. The MIRI spectrum is shown in black. The different molecular emissions are shown using slab model spectra as colored filled regions in each panel. `?' indicates more modeling is required to confirm the detection.}
    \label{fig:j1605ch3}
\end{figure}

The VLMS spectra are typically show rich molecular emission as can be seen in Fig.\,\ref{fig:features_summary}. In all of our sources, the SNR is above 100 in the spectral windows where molecular emissions are observed (more details on the SNR calculation from the ETC is presented in Jang et al. in prep.). Most of the molecular emission overlap in their wavelengths, and their detectability is influenced more by the surrounding molecular emission rather than the noise. i.e., defining a simple criteria classifying a $Q$-branch peak of a species above 3\,$\sigma$ as a detection, similar to the case of T\,Tauri disks, is not complete. The detection criteria should also account for neighboring molecular emission. This requires fitting the spectra with slab models, which is beyond the scope of this work. Instead we include all the detections reported by the previous publications on individual sources:
\begin{itemize}
    \item J0438: \citet{2025arXiv250411424P}
    \item NC1: Morales-Calder\`on et al. (subm.)
    \item IC147: \citet{2024Sci...384.1086A}
    \item Sz28: \citet{2024A&A...689A.231K,2024A&A...690A.100K}
    \item TWA27: Patapis et al. (subm.)
    \item J1605: \citet{2023NatAs...7..805T}, Kanwar et al. (subm.)
\end{itemize}
For the remaining sources (NC9, J0439, HKCha, and J1558) we refer to \citet{2025ApJ...984L..62A} for discussion on detections of \ch{H_2O}, \ch{OH}, and \ch{CO}. The rest of the atomic and molecular detections are illustrated in Figs.\,\ref{fig:j1558Hydrogen}-\ref{fig:nc9_c2h6}. \hl{The synthetic molecular spectra in these figures and generated from slab models introduced in \citet{2024Sci...384.1086A}.} In the J1558 spectrum (bottom left panel of Fig.\,\ref{fig:remaindetect}), the \ch{CO_2} model $Q$-branch peak matches with one of the observed peaks but the shape does not - the broader shoulder on the short wavelength side of the peak is missing in the observed spectrum. There is also a pure rotational water line (19$_{9,10}$-18$_{8,11}$) with high Einstein A coefficient that matches the peak. Proper modeling of the source is required to confirm the presence of \ch{CO_2}. In the panels on the right, while some peaks match with the observed spectrum of HKCha and J0439, some at $\sim$16.6\,$\mu$m do not. This can be due to several factors such as different continuum level, excitation conditions, or simply absence of \ch{CH_3}. Again, proper modeling of the spectra is required for these sources. The top and middle panels of Fig.\,\ref{fig:nc9_c2h6} shows hints of \ch{CH_4} and \ch{C_2H_4}, and the bottom panel shows firm detection of \ch{C_2H_6}. In addition to previous detections in J1605, we detect \ch{CH_3} (Fig.\,\ref{fig:j1605ch3}).

\FloatBarrier

\section{Detection rate correlations}
\begin{figure*}[h!]
    \centering
    \includegraphics[trim={0 0 0 1em}, clip, width=\linewidth]{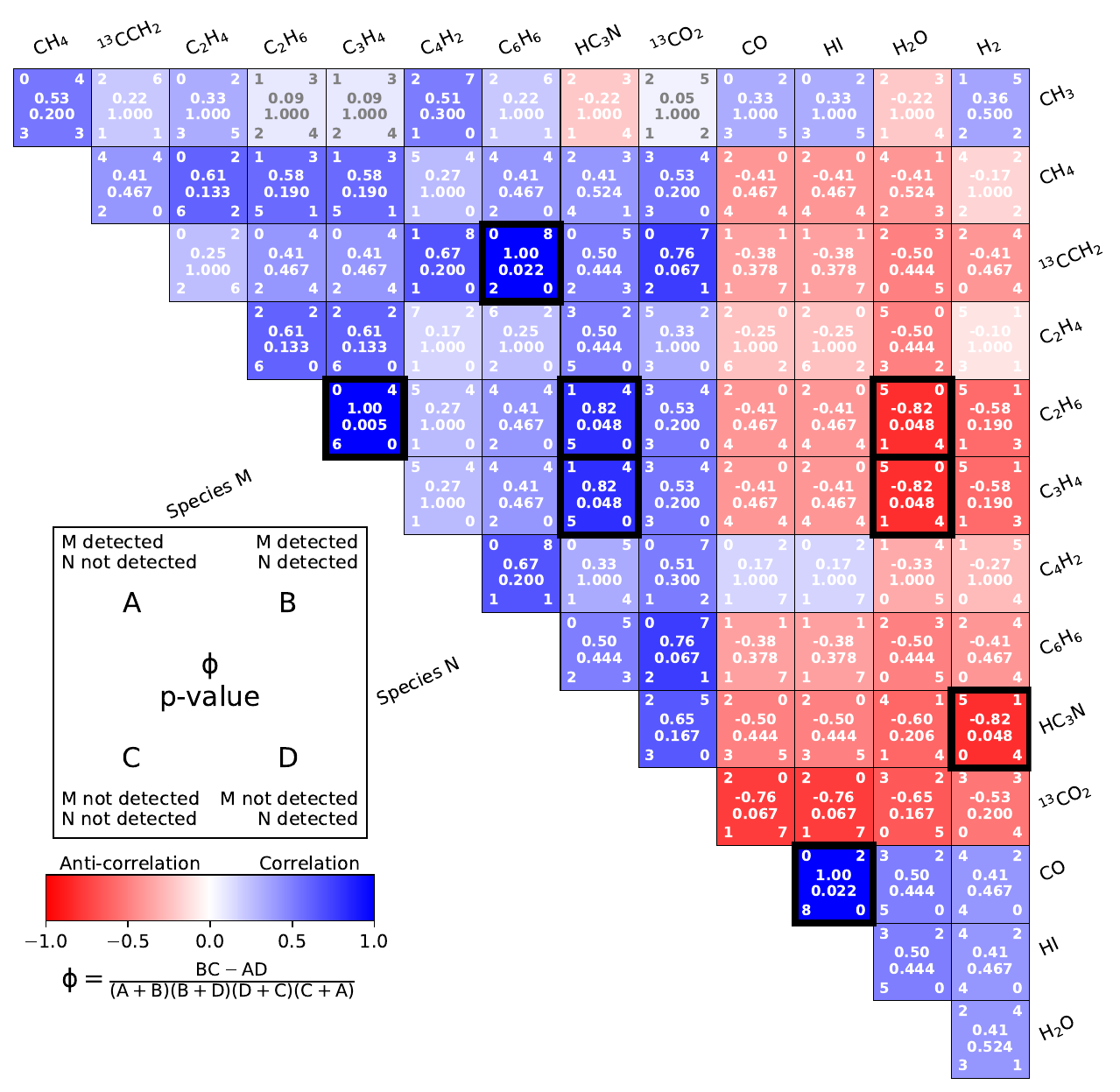}
    \caption{Binary correlation coefficients \citep{Yule1912} between the detection rates of different molecules. The box on the bottom left describes the details of various numbers shown in each colored box on the right. The color denotes the value of the correlation coefficient, with red corresponding to an anti-correlation and blue corresponding to a correlation. For calculating the coefficients only firm detections are considered as `detected' while the rest (including tentative detections) are considered `not detected'. Molecules that are detected in all the sources or are detected in only one source are not shown. Molecule pairs with statistically significant correlations (p-value<0.05) are highlighted with black borders.}
    \label{fig:correlations}
\end{figure*}

Figure\,\ref{fig:correlations} shows the binary correlation coefficients for the molecules detected in the spectra. Since the species detected in all sources and those detected in only one source do not provide useful correlations, those species are not shown. The correlations and anti-correlations are indicated by colors varying from blue to red. Interestingly, Fig.\,\ref{fig:correlations} shows three distinct regions: i) the blue-dominated region to the left of the figure, which largely indicates correlations between organic molecule pairs; ii) the smaller blue region at the bottom, which shows correlations between the inorganic species; and iii) the red region on the right that shows the anti-correlation between the organic molecules and inorganic species. We also calculate the p-value using the Fisher's exact test (\texttt{scipy.stats.fisher\_exact}). We find statistically significant strong correlations between less abundant hydrocarbons, and statistically significant strong anti-correlations between less abundant hydrocarbons and inorganic molecules.

\end{document}